\documentclass[5p]{elsarticle}

\usepackage[utf8]{inputenc}
\usepackage[per-mode=symbol-or-fraction]{siunitx}
\DeclareSIUnit\micron{\micro\metre}
\DeclareSIUnit\Gbits{\giga\bit\per\second}
\DeclareSIUnit\mWsqcm{\milli\watt\per\cm\squared}
\DeclareSIUnit\gps{\gram\per\second}
\DeclareSIUnit\kCHF{\kilo CHF}
\DeclareSIUnit\mbar{\milli\bar}
\DeclareSIUnit\rpm{rpm}
\usepackage{booktabs}
\usepackage{chemformula}
\usepackage{hyperref}
\usepackage{multirow}
\usepackage[stable]{footmisc}

\usepackage{caption}
\newcommand{\mm}[1]{\qty{#1}{\milli\meter}}
\newcommand{\mwcm}[1]{\qty{#1}{\milli\watt/\centi\meter^2}}

\newcommand{\encirc}[1]{\raisebox{.5pt}{\textcircled{\raisebox{-.9pt} {#1}}}}

\newcommand{\qty}[2]{\SI{#1}{#2}}
\newcommand{\unit}[1]{\SI{}{#1}}
\newcommand{\qtyrange}[3]{\SIrange{#1}{#2}{#3}}

\title{Successful cooling of a pixel tracker using gaseous helium: \\studies with a mock-up and a detector prototype.}

\author[1]{Thomas Theodor Rudzki\corref{cor1}} \ead{rudzki@physi.uni-heidelberg.de}
\author[2]{Frank Meier Aeschbacher} \ead{frank.meier@psi.ch}
\author[3]{Marin Deflorin} \ead{marin.deflorin@fhnw.ch}
\author[3]{Niculin Flucher} \ead{niculin.flucher@fhnw.ch}

\cortext[cor1]{Corresponding author: \url{rudzki@physi.uni-heidelberg.de}}  

\address[1]{Physikalisches Institut, Universit\"at Heidelberg, Im Neuenheimer Feld~226, 69120~Heidelberg, Germany}
\address[2]{Laboratory for Particle Physics, Paul Scherrer Institut, Forschungsstrasse~111, 5232~Villigen, Switzerland}
\address[3]{Institute of Thermal and Fluid Engineering, FHNW University of Applied Sciences and Arts Northwestern Switzerland, Klosterzelgstrasse~2, 5210~Windisch, Switzerland}

\begin{document}\sloppy

\begin{abstract}
We report the successful operation of a functional pixel detector with gaseous helium cooling. Using an accurate mock-up beforehand, the cooling was validated. We use a miniature turbo compressor to propel the helium at \qty{2}{\gps} under ambient pressure conditions with gas temperatures above \qty{0}{\celsius}. Our earlier results based on computational fluid dynamics simulations and a much simpler mock-up are confirmed. With this, we paved the path to cool pixel detectors in experimental particle physics at heat densities up to \qty{400}{\mWsqcm} using helium. This enables cooling of detectors with very low mass requirements, minimising the effects of multiple Coulomb scattering effectively. The concept presented here is not limited to pixel detector applications and can be used to cool any surface with comparable heat-densities, only limited by shaping the helium gas flow.

\end{abstract}

\maketitle

\tableofcontents

\section{Introduction}

Silicon pixel detectors became a staple in particle physics for building tracking detectors. Their superb position resolution in the few \qty{10}{\micron} range and high track rate capabilities are among the reasons for this success. However, being active electronic components, they dissipate heat while operated. In the densely packed tracking detectors of modern particle physics experiments, the heat needs to be actively removed from the detectors to keep their performance under control. Furthermore, operation at temperatures well below \qty{0}{\celsius} is sometimes required to limit effects of radiation damage when used in high dose environments.

From the beginning, different approaches for cooling have been implemented.
For example when looking at LHC experiments, CMS used mono-phase liquid cooling with \ch{C6F14} as coolant for the pixel detector of LHC Run~1. 
Their barrel alone produced about \qty{2.6}{\kilo\watt} of heat~\cite{CMS:2008xjf}. 
In contrast, ATLAS and ALICE already used a bi-phase cooling system~\cite{ATLAS:2008xda,ALICE:2008ngc}.
ATLAS had a barrel detector dissipating up to \qty{10}{\kilo\watt}, hence the bi-phase approach provided the necessary cooling power by using the phase transition from liquid to vapour. Both used short-chained fully fluorinated hydrocarbons. The choice of these coolants have the benefit of being non-conductive and they have a high vapour pressure. Both properties help mitigating risks of detector damage in case of leaks in the cooling circuits. Some systems received upgrades, e.g.~CMS switched to \ch{CO2} bi-phase cooling for their phase~1 detector with a much increased instrumented surface. ALICE took the advantage of their new low-power pixel detector to switch to a liquid water cooling system. As a smart feature, they operate the circuit with a pressure well below ambient. Any small leak will lead to a detectable air intake and not a spill of liquid coolant. A synopsis of barrel-shaped pixel detectors is given in \autoref{tab:pixeldetscomparison}.

However, all these systems require tubing, appropriately chosen and laid out to the needs of the experiments.
This adds dead material to the tracking detectors. 
For applications with particle tracks at sufficiently high momenta (e.g. greater than a few \qty{100}{\MeV}), the amount of multiple Coulomb scattering may be acceptable. 
For applications with low momentum tracks, e.g.~in precision physics or heavy ion colliding experiments, the scattering induced  distortion will lower the track reconstruction precision or will make the experiment impossible to pursue. 
Gaseous detectors like wire chambers provide a viable option with their inherently low-mass properties. 
Their constrained particle rates limit the physics reach of such experiments.
With recent developments in very thin pixel detectors, a new class of experiments was made possible but at the expense of the need for novel cooling concepts: ALICE uses them to tackle the detector challenges of the unique low-momenta character of most of the tracks in heavy-ion collisions~\cite{Blidaru:2021kdy}. Mu3e re-ignites the search for $\mu^+ \rightarrow e^+e^-e^+$ after decades of no activity on this decay~\cite{Mu3e:2020gyw}. Also noteworthy is the STAR detector, where monolithic active pixel sensors (MAPS) were used, actively cooled by air under ambient conditions~\cite{Greiner:2011zz,Schambach:2014uaa}. 
Another approach is pursued by the Belle~II pixel detector, where a forced nitrogen flow along the pixel modules complements the combined bi-phase \ch{CO2} cooling. Their design also cools the active readout electronics and the ends of the modules. The pixel chip itself is a unique DEPFET design with very low heat dissipation in the active area. 

Here, we report on the cooling approach Mu3e follows. 
This pixel detector, currently under construction, has a comparable heat density and instrumented surface to many of the existing pixel barrel detectors (see \autoref{tab:pixeldetscomparison}\footnote{Throughout this paper, we state different heat densities. \qty{250}{\mWsqcm} is the expected nominal heat dissipation from the chip and the losses in the HDI. \qty{350}{\mWsqcm} is the maximum allowed heat dissipation  of the chip the cooling system has to be designed for. \qty{400}{\mWsqcm} adds contingency, including losses in HDI and connectors. This offers a headroom of almost a factor of 2. The observed heat density with the latest versions of the \textsc{MuPix} chip family is below the value given in the table (\qty{250}{\mWsqcm}), see \autoref{sec:VtxPrototype}.}). 
To reduce the material in the pixel detector even more, the Mu3e experiment uses MAPS-based \textsc{MuPix} chips~\cite[Chap.~8]{Mu3e:2020gyw} on very thin aluminium-polyimide high density interconnects~\cite[Sec.~7.2.5]{Mu3e:2020gyw}.
In addition, helium gas is used as coolant~\cite[Sec.~12.2]{Mu3e:2020gyw}. 
The gas is at ambient temperatures and pressures.
The low density and atomic mass of helium reduces the multiple Coulomb scattering of the particles passing through space.
However, precisely these properties make it particularly difficult to feed helium, either using a pump or a compressor. 
Such standard equipment is mostly designed for use with air. 
The engineer's approach is usually to go for a higher density by increasing the helium pressure to $\mathcal{O}(\qty{10}{\bar})$. 
This is a very energy intensive process. For example, oil-lubricated screw compressors used for helium liquefaction plants would be a standard solution, but such a compressor requires about \qty{250}{\kilo\watt} of electrical energy\footnote{This power rating is taken from typical compressors used in industry-standard helium liquefaction plants.} to pump the \qty{55}{\gps} of helium, the required mass flow Mu3e would need to cool its detector. 
Hence a novel approach has been developed based on miniature turbo compressors that became commercially available recently. This paper will report on the results achieved with this setup, proving the feasibility of using gaseous helium as an effective coolant for pixel tracking detectors.

\begin{table*}[h!]
    \centering
    \caption{Overview of cooling parameters of various pixel barrel detectors. In cases where the pixel detector also has forward disks, values are scaled by area. 
    See~\ref{app:pixel_barrel_table} for details about the values used for calculating the instrumented areas and the total power.}
    \label{tab:pixeldetscomparison}
{\footnotesize
    \begin{tabular}{llccccccl}
            \toprule
        \multicolumn{2}{l}{Experiment} & Coolant   & Phase     & Target Temp. & Heat density & Instr. area & Total power & Ref \\
                   &            &           &           & \unit{\celsius} & \unit{\mWsqcm} & \unit{\m\squared} & \unit{\watt} & \\
            \midrule
        CMS & LHC Run 1         & \ch{C6F18}& liq       & -10 & 333 & 0.78 &  2600 & \cite{CMS:2008xjf} \\ 
            & Phase 1 upgrade   & \ch{CO2}  & liq/vap   & -20 & 500 & 1.20 &  6000 & \cite{CMSTrackerGroup:2020edz} \\
            \midrule
        ATLAS & LHC Run 1       & \ch{C3F8} & liq/vap   &  -7 & 444 & 2.25 & 10000 & \cite{ATLAS:2008xda} \\
            \midrule
        ALICE & LHC Run 1       & \ch{C4F10}& liq/vap   & +25 & 643 & 0.21 &  1350 & \cite{ALICE:2008ngc} \\
              & Upgrade IB      & \ch{H2O}  & liq       & +25 & 300 & 0.19 &   570 & \cite{ALICE:2013nwm} \\
              & Upgrade OB      & \ch{H2O}  & liq       & +25 & 100 & 10.7 &  10700 & ibid. \\
            \midrule
        STAR  &                 & air       & gaseous   & +25 & 170 & 0.16 &   272 & \cite{Greiner:2011zz,Schambach:2014uaa} \\
            \midrule
        Belle II & PXD          & \ch{N2} + \ch{CO2}   & gaseous + liq/vap  & +25 & 182 & 0.033 & 60  & \cite{Ye:2016rwb,Belle-IIDEPFET:2021pib} \\
            \midrule
        Mu3e  & Vertex          & \ch{He}   & gaseous   &   0 & 250 & 0.052&   130 & \cite{Mu3e:2020gyw} \\
              & Outer layers    & \ch{He}   & gaseous   &   0 & 250 & 1.31 &  3276 & ibid. \\
            \bottomrule
    \end{tabular}
    }
\end{table*}

\section{Methods}

\subsection{Vertex detector description and test setups}
\begin{figure}[h!]
	\centering
	\includegraphics[width=0.95\columnwidth]{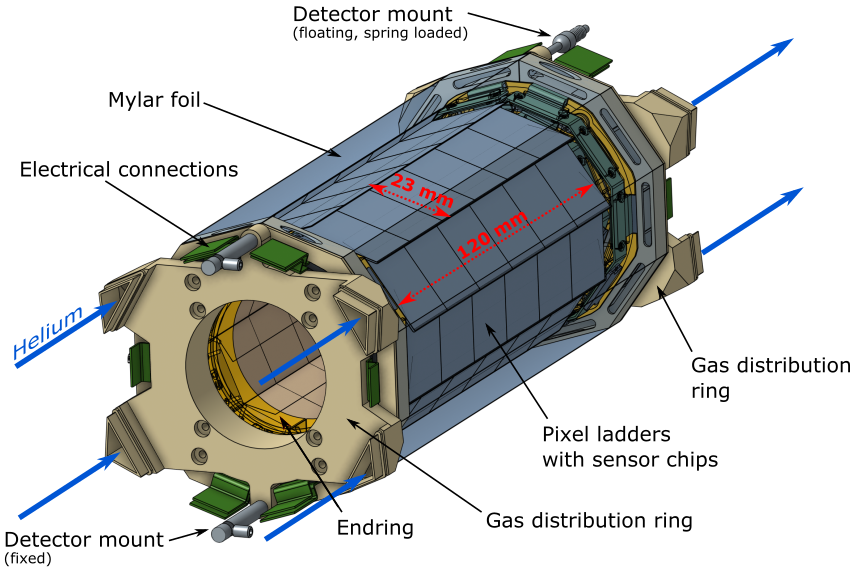}
	\caption{Schematic view of the vertex pixel barrel. Each of the 8+10 ladders carries 6 sensor chips. End-rings provide the mechanical support to the ladders. \cite{Mu3e:2020gyw}. The inner layer is only visible through the opening in the end-ring. The mylar foil has been trimmed in this view for better visibility and covers the full barrel in reality.
	}
	\label{fig:inner_tracker1}
\end{figure}

The innermost part of the Mu3e pixel detector, subsequently called the \textit{Mu3e vertex detector}, is a barrel made of two pixel detector layers, where the active part is approximately \qty{12}{\cm} long, see~\autoref{fig:inner_tracker1}.
The inner layer consists of 8 ladders, the outer of 10 ladders. Each ladder carries 6 sensors in a row.
The mechanical structure of a ladder is a high-density interconnect (HDI) made of a custom aluminium-polyimide laminate. 
It provides all electrical connections for power, sensor bias voltage, and differential signal lines for control and readout.
The chips are glued onto the HDIs.
The electrical connections between the HDI and the chip are made using \emph{single point tape automated bonding} (spTAB)~\cite{Lau1992} and~\cite[Sec.~3]{MeierAeschbacher:2020ldo}.\footnote{A variant of \emph{tab automated bonding} is used where two-layer aluminium-polyimide ribbons are directly bonded to pads on a chip or a PCB. Ultrasonic bonding is used to form the contact. In this use case, the ribbons are the traces on the HDI in an opening of the polyimide insulating layer. No extra material as the wire in wire bonding is needed.}
The gaseous helium is distributed in two flows by a 3D printed gas distribution ring, see \autoref{fig:inner_tracker1}).
The first flow is between the two layers, and the second around the outer layer, confined by a thin mylar foil surrounding the detector.

Dedicated to the cooling studies, two mock-ups of the vertex detector at different levels of accuracy in matching the final detector have been constructed.
The difference is the choice of materials, as explained below.
Both of them can be heated with heat densities up to \SI{400}{\mWsqcm}. This matches the performance specification of the cooling system, as required by the Mu3e pixel detector project.
In addition, a functional vertex detector prototype with working pixel chips in a modified geometry has been operated with gaseous helium as coolant.

The first mock-up studies were based on simple, readily available materials, and led to important results, see~\cite{MeierAeschbacher:2020ldo,Deflorin:thesis}. 
This led to an improved and accurate \emph{silicon heater mock-up} and a working \emph{pixel detector prototype}, described below.

\subsubsection{Silicon heater mock-up}
\begin{figure}[t!]
	\centering
	\includegraphics[width=0.8\columnwidth]{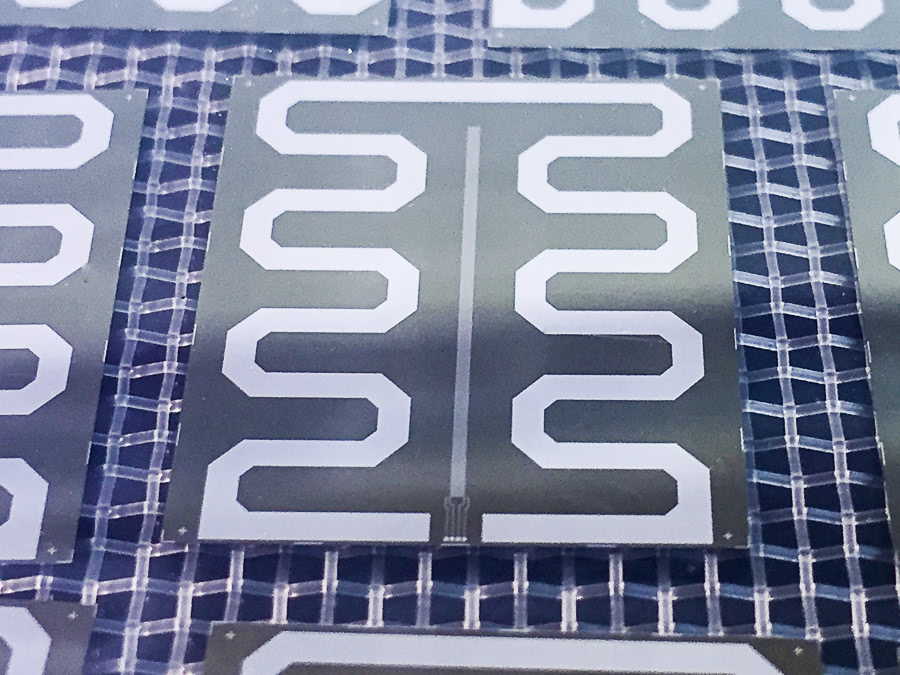}
	\caption{Silicon heater chip~\cite{Mu3e:2020gyw}. The large, readily visible meander is for heating. The vertical bar in the centre with visible bond-pads is a thinly-structured meander for temperature sensing.
	}
	\label{fig:silicon_heater}
\end{figure}

The silicon heater mock-up matches accurately both the geometry and the materials of the final vertex detector.
It consists of \qty{50}{\micron} thin silicon heater chips\footnote{Fabricated by MPG Halbleiterlabor in Munich to our designs.} and HDIs\footnote{Manufactured by LTU, Kharkiv, Ukraine, \url{http://www.ltu.ua/}}.
The chips feature a heating meander ($R \approx \qty{2.75}{\ohm}$) and a meander for resistive temperature measurement ($R \approx \qty{1250}{\ohm}$) which were photolithographically etched from an aluminium layer sputtered on the silicon wafer (\autoref{fig:silicon_heater}).
The HDIs provide individual power for the heating of the chips and readout lines for 4-wire temperature sensing on chip.
A more detailed description is given in \cite[Sec.~7.5]{Mu3e:2020gyw} and \cite{Rudzki:2021smh}.

In contrast to the \textsc{MuPix} chips, which dissipate most heat in their periphery\footnote{Non-sensing edge of the pixel chip with active electronics (amplification, communication, power regulation etc.) and bond pads.}, the silicon heater chips are heated uniformly by the meander.
Therefore, the cooling studies give information about the average temperatures of each chip but not about the temperature peaks expected in the peripheries.
The maximum allowed chip temperature of \SI{70}{\degreeCelsius} is given by the glass transition temperature of the adhesives used.
The glue joints between chip and HDIs are located at the active pixel matrix, which is why hot spots in the periphery play only a marginal role for the mechanical stability.

\subsubsection{MuPix10 vertex detector prototype}
\begin{figure}[h!]
	\centering
	\includegraphics[width=\columnwidth]{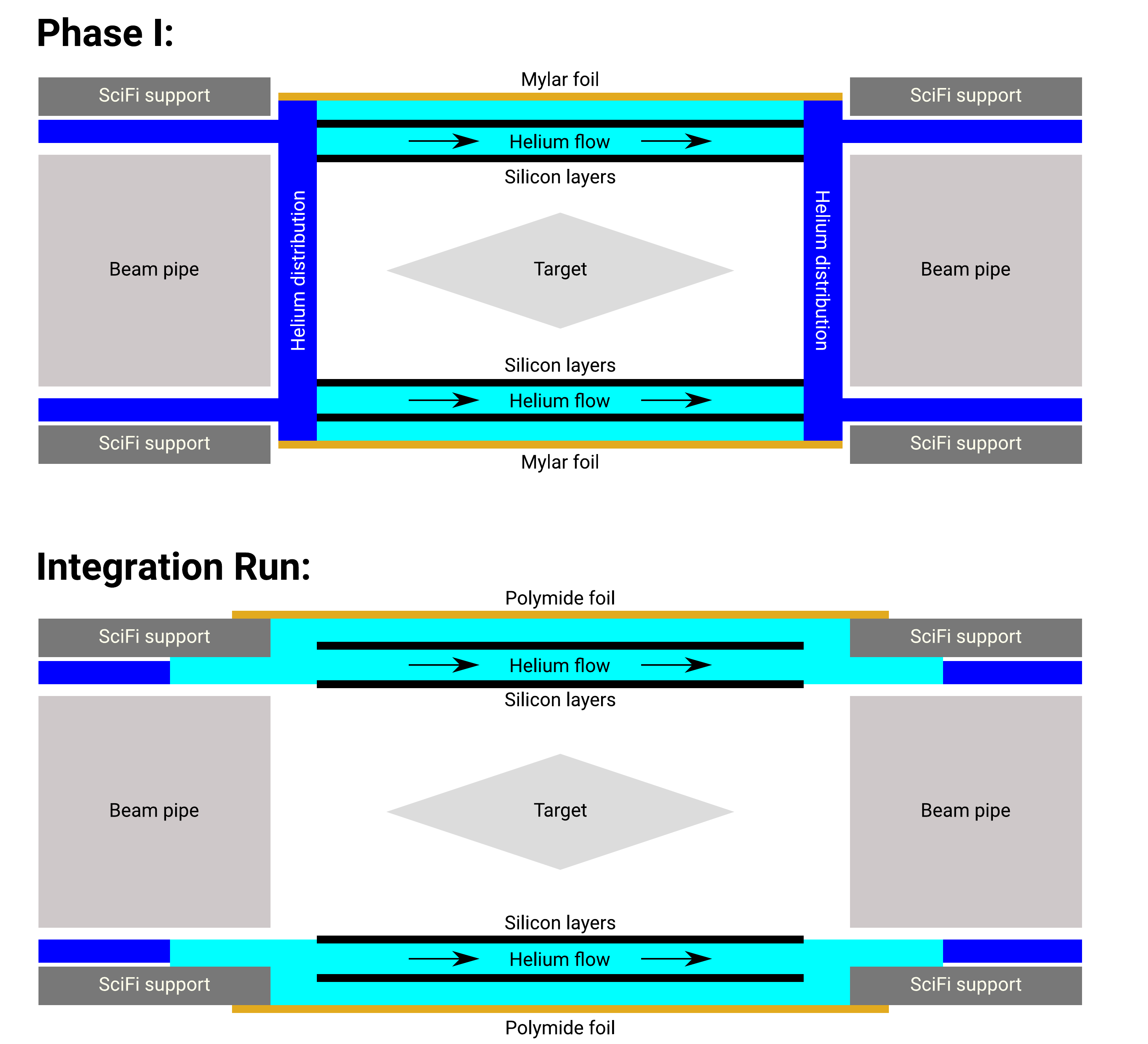}
	\caption{Sketch of the helium flow distribution for the final vertex detector (phase I) and the DAQ integration run configuration~\cite{Rudzki:thesis}.
	}
	\label{fig:he_vertex_prototype}
\end{figure}

This is a fully working prototype using \textsc{MuPix10} MAPS chips~\cite{Augustin:2020pkv}.
The detector was successfully operated to observe cosmic ray muons~\cite{koeppel_proceedings} as well as with muons from the Paul Scherrer Institut (PSI) secondary beamline $\pi$E5~\cite{Koppel:2022kbd}.
The campaigns focused on the demonstration of the full operation of the vertex detector, thermal studies were of lower priority. 
For simplicity, conventional PCBs have been used instead of an HDI, which changes the thermal properties compared to the final detector design. 
The minimum radii of the two pixel layers increased from \mm{23.3}/\mm{29.8} to \mm{31.7}/\mm{44.6}.
The helium distribution also received changes as depicted in \autoref{fig:he_vertex_prototype} since the gas distribution ring of the final detector is not integrated.
In addition, the PCBs differ thermally from the HDIs intended for the final detector. 
They have a much larger thermal mass and allow for a notable conductive heat transport, acting as a heat-sink. 
Thus, comparisons of temperature distributions obtained by simulations and the mock-ups mentioned above will be of limited value.

\begin{figure}[h!]
	\centering
	\includegraphics[width=\columnwidth]{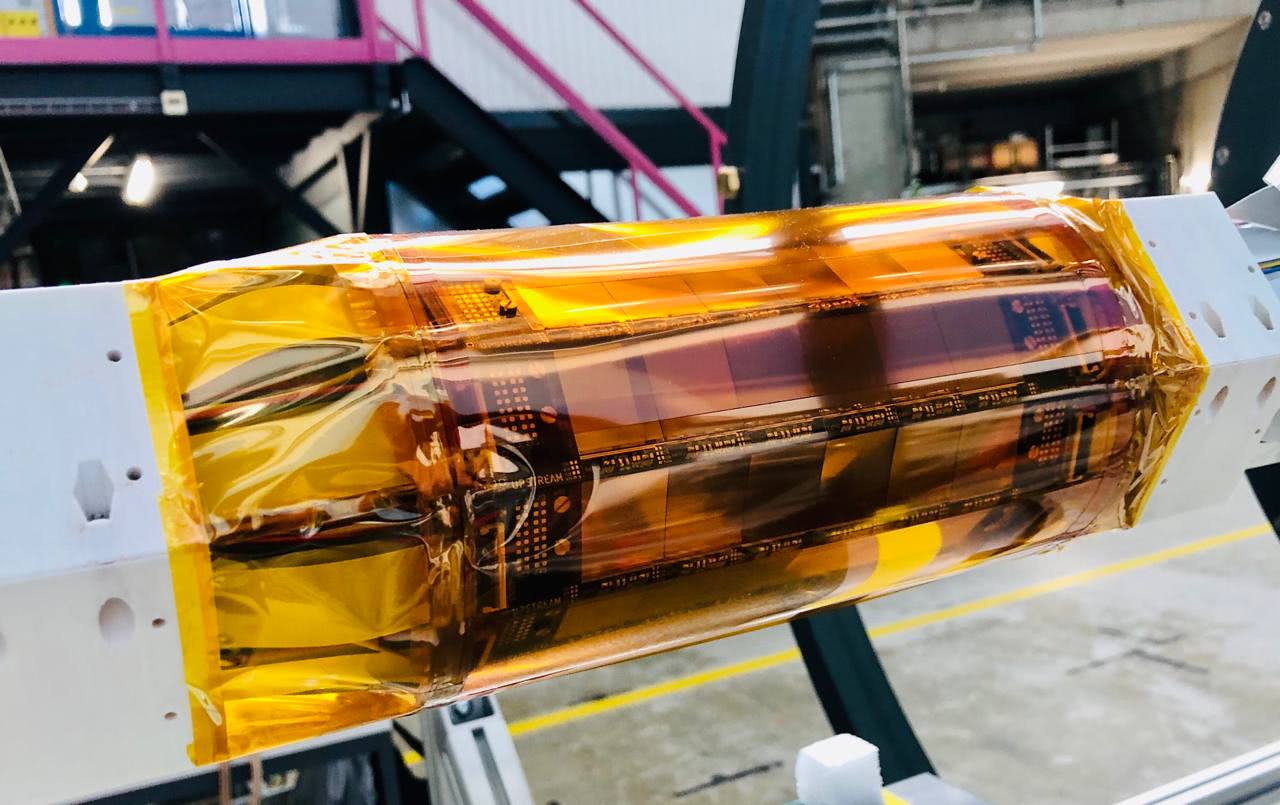}
	\caption{\textsc{MuPix10} vertex detector prototype with a polyimide foil confining the outer helium flow. The 3D~printed support (white) of the Mu3e SciFi detector serves as well as helium confinement, tubes supplying helium end beneath.~\cite{Rudzki:thesis}.
	}
	\label{fig:vertex_prototype}
\end{figure}

\subsection{Cooling infrastructure} 
\label{cooling_setup}

\begin{figure}[h!]
	\centering
	\includegraphics[width=\columnwidth]{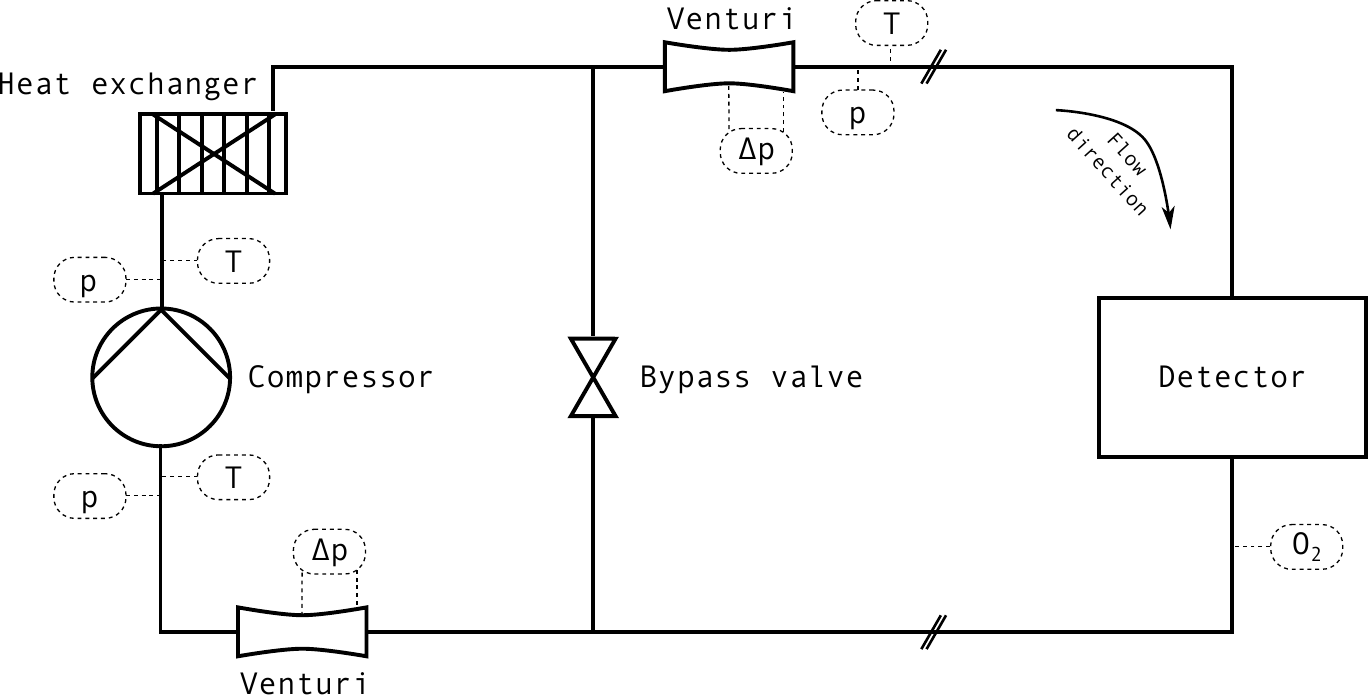}
	\caption{Conceptual process flow diagram of the helium cooling system. Key to sensors: $T$ temperature, $p$ absolute pressure, $\Delta p$ differential pressure, $\ch{O2}$ oxygen concentration.
	}
	\label{fig:HePlantPFDsimple}
\end{figure}

\begin{figure}[h!]
	\centering
	\includegraphics[width=\columnwidth]{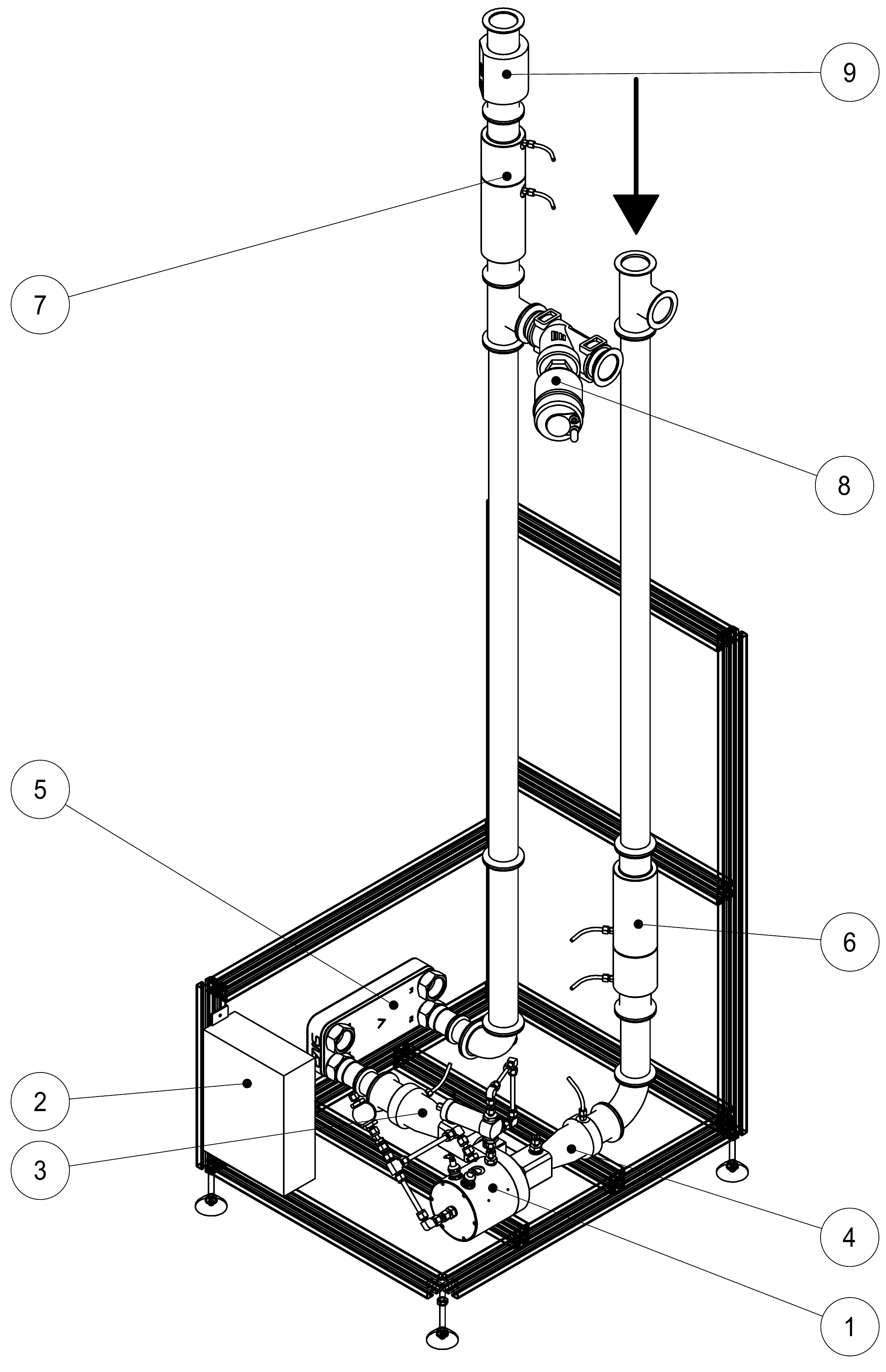}
	\caption{Rack for miniature helium compressor. \encirc{1} Miniature turbo-compressor, \encirc{2} power converter for the compressor, \encirc{3} diffusor and \encirc{4} confusor with integrated temperature and pressure sensors, \encirc{5} heat exchanger (cooling water hook-up not shown), \encirc{6} and \encirc{7} Venturi tubes, \encirc{8} bypass valve (connected using a corrugated steel hose to the T piece in the right, not shown), \encirc{9} temperature and pressure sensors at the outlet. Some structural elements of the rack are removed for better visibility. The footprint of the rack is about $\qty{1}{\metre} \times \qty{1}{\metre}$.
	}
	\label{fig:rack_comp}
\end{figure}

The cooling of the vertex detector with gaseous helium requires a flow around \SI{2}{\gps} at ambient pressure~\cite{Deflorin:thesis}. 
\autoref{fig:HePlantPFDsimple} shows the schematic of the helium cooling setup.
A miniature turbo compressor is used for feeding the helium at the required mass flows, housed in a rack carrying all the other equipment needed for operation, shown in \autoref{fig:rack_comp}. 
The whole system is designed using standard ISO-KF small flange components in DN~50 size. 
For the compressor, ISO-KF flanges are not available from the suppliers.
Custom adaptors made in-house are in use.
What follows is a description of the components.

\subsubsection{Miniature turbo compressor}

A Celeroton model CT-NG-2000 miniature turbo compressor\footnote{Celeroton AG, Volketswil, Switzerland, \url{https://www.celeroton.com/}.} is used to pump the helium at the required mass flows. The manufacturer modified the design to accommodate for helium at ambient pressures.
The compressor uses an impeller with gas-bearings and a fully integrated electric drive to reach speeds of up to \qty{240}{\kilo\rpm}, controlled by a custom power converter supplied by the manufacturer.
While optimised for helium under ambient conditions, operation with other gases like air is allowed at a reduced rotation speed, resulting in a decrease of the overall performance.
The compressor is relatively small, its housing being approx.~$\qty{200}{\mm}\times\qty{160}{\mm}\times\qty{130}{\mm}$, its shaft power around \qty{150}{\watt} at nominal conditions for propelling \qty{2}{\gps} of helium.

\begin{figure}[h!]
	\centering
	\includegraphics[width=\columnwidth]{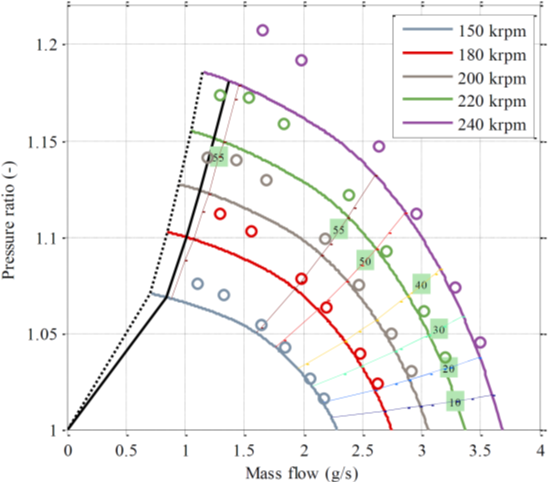}
	\caption{Compressor map measured by Celeroton with helium. The black dotted line is the surge limit, the solid line has an added safety factor. Operation in the area left of the surge line would lead to potential damage. \emph{Map courtesy Celeroton, used by permission.}
	}
	\label{fig:map_comp}
\end{figure}

Operation of a turbo compressor requires careful monitoring of the mass flow $\dot{m}$ and the pressure ratio $\Pi=p_\text{out}/p_\text{in}$.
If the compressor surpasses the stability limit, rotating stall or surge will occur at some point and damage the impeller.
The compressor map in \autoref{fig:map_comp} is measured with helium. The maximum achievable pressure ratio is therefore around \qty{1.22}{} which results in a mass flow of \qty{1.6}{\gps}. The maximum useful mass flow is \qty{3.5}{\gps} at a lower pressure ratio of about \qty{1.05}{}. The required mass flow of \qty{2.0}{\gps} is well covered with this compressor at pressure ratios even above \qty{1.15}{}.

\paragraph{Dimensionless compressor map}
To become independent of the media, a dimensionless compressor map is beneficial.
To obtain this, a dimensionless flow $\Phi$ and a pressure coefficient $\Psi$ are required. The circumferential Mach number $Mu_2$ at the outer diameter of the impeller must be constant for different media to ensure similar flow conditions. Therefore, the rotational speed must be adjusted depending on the media. The flow coefficient $\Phi$ is computed using the mass flow $\dot{m}$, the static gas density at the inlet $\rho_1$, the outer circumferential speed $u_2 = \frac{2 \cdot \pi \cdot n \cdot D_2}{2 \cdot 60}$ where $n$ is the rotational speed and $D_2$ the outer diameter of the impeller. All together gives
	\begin{equation}
		\Phi = \frac{4 \cdot \dot{m}}{\pi \cdot \rho_1 \cdot u_2 \cdot D_2^2}
		\label{eqn:Phi}
	\end{equation}
The pressure coefficient $\Psi$ is computed using the constant heat capacity $c_p$, the total temperature at the inlet $T_1^0$, the pressure ratio $\Pi$, the heat capacity ratio $\kappa$ and the circumferential speed $u_2$ from above:
	\begin{equation}
		\Psi = \frac{c_p \cdot T_1^0 \cdot \Pi^{(\frac{\kappa-1}{\kappa})}}{u_2^2}
		\label{eqn:Psi}
	\end{equation}
For the circumferential Mach number, the specific gas constant is additionally required to estimate the speed of sound:
	\begin{equation}
		Mu_2 = \frac{u_2}{\sqrt{\kappa \cdot R_s \cdot T_1^0}}
		\label{eqn:Mu2}
	\end{equation}

\begin{figure}[h!]
	\centering
	\includegraphics[width=\columnwidth]{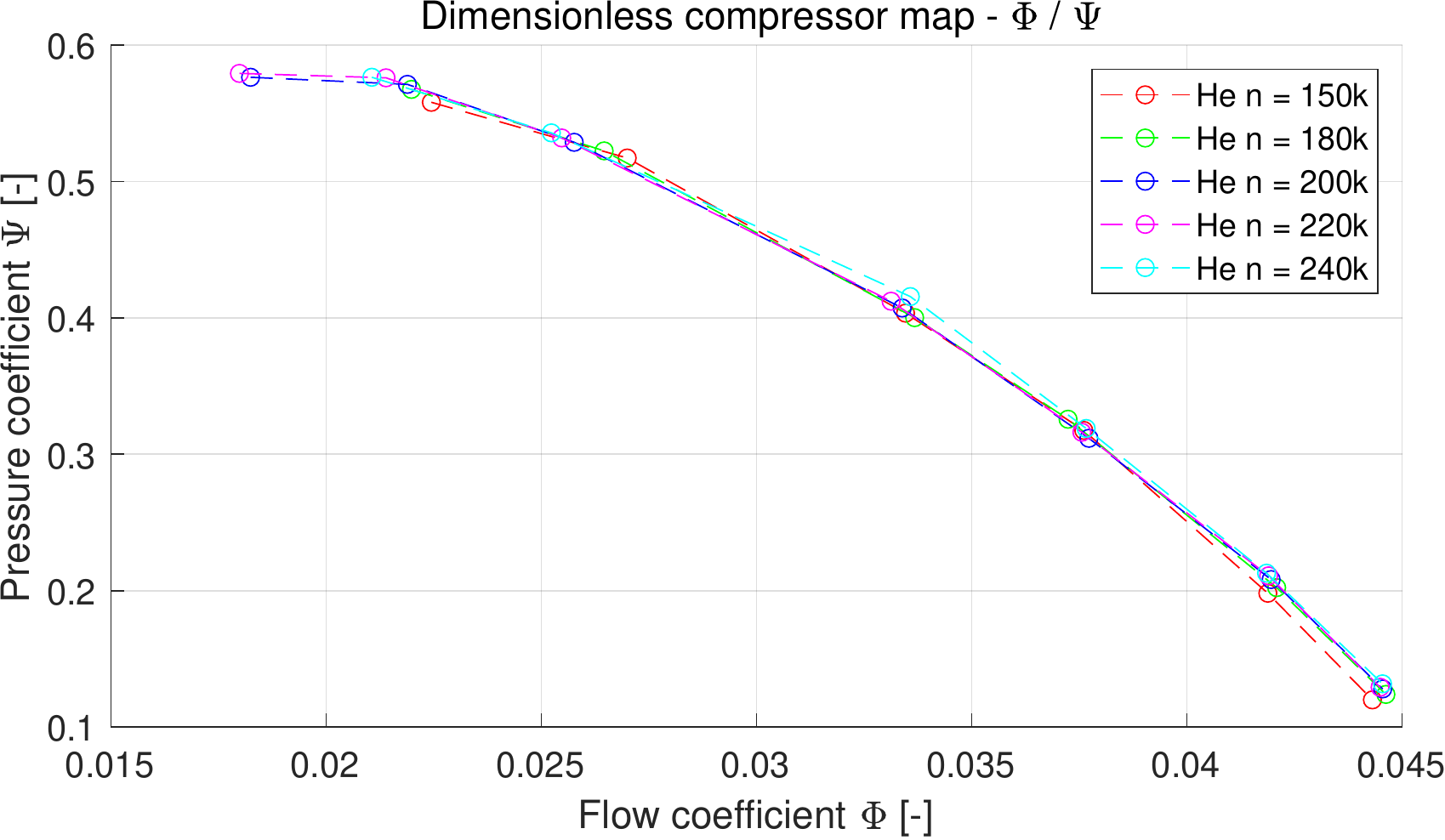}
	\includegraphics[width=\columnwidth]{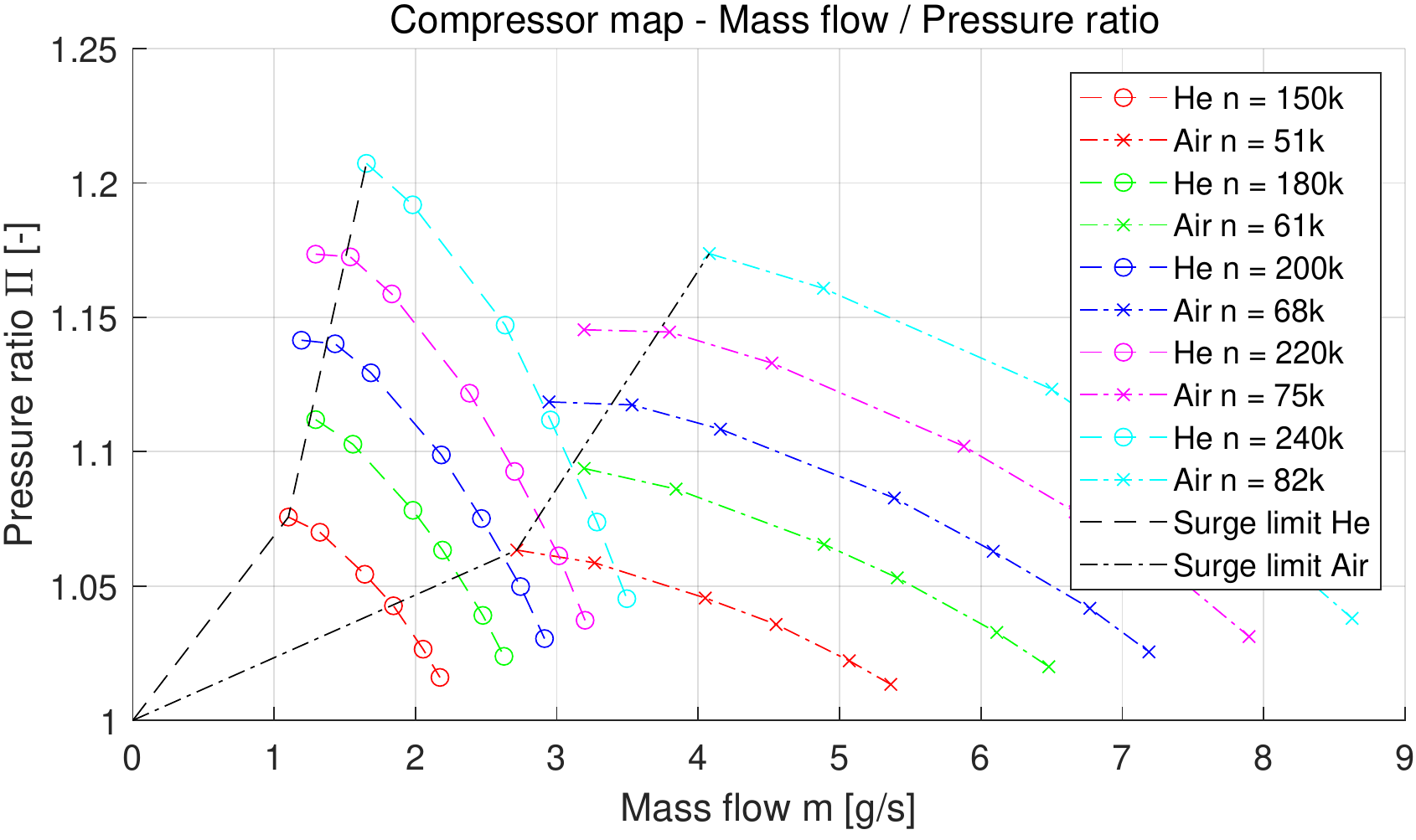}
	\includegraphics[width=\columnwidth]{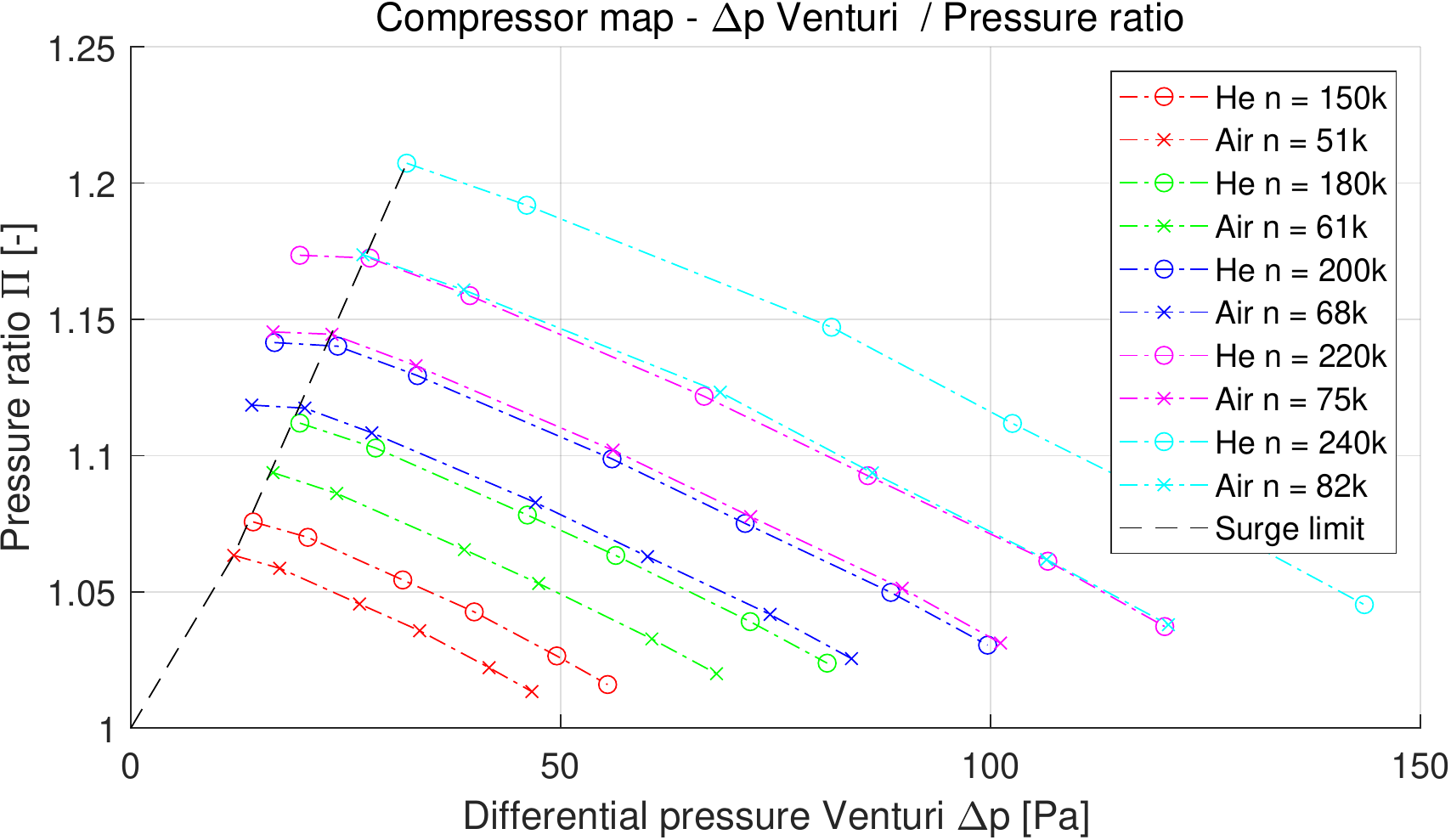}
	
	\caption{Dimensionless compressor map based on the measurement data from \autoref{fig:map_comp}. Top: Dimensionless compressor map. Middle: Compressor map computed for air. Bottom: Compressor map with the differential pressure at the Venturi tube on the x-axis.
	}
	\label{fig:map_comp_dim}
\end{figure}

Based on \autoref{eqn:Phi} and \autoref{eqn:Psi}, the compressor map measured with helium (\autoref{fig:map_comp}) was converted into a dimensionless map (\autoref{fig:map_comp_dim} top), where all the rotational speed curves overlap. Therefore, the compressor shows similar behaviour in the range between \qtyrange{150}{250}{\kilo\rpm}. Using \autoref{eqn:Mu2}, the rotational speed for the compressor can be estimated for another medium, e.g.~air, and the compressor map can be generated (\autoref{fig:map_comp_dim} centre). The range for the operation with air is between \qtyrange{51}{82}{\kilo\rpm}. In terms of the mass flow, there is more air being compressed by the compressor but as the density of helium is ten times lower than air, the volume flow is higher with helium and also the pressure ratio. The online monitoring of the surge limit is then dependent on the air/helium concentration, which is more complex. To simplify the surge monitoring, a map was created where the abscissa is switched to the differential pressure of the Venturi tube (see \autoref{sub:Venturi}) instead of the mass flow. This map is shown in \autoref{fig:map_comp_dim} bottom and for both helium and air the map is quite similar at the same circumferential Mach number. This has the advantage, that the surge limit can be monitored only by the differential pressure at the Venturi tube and the pressure ratio over the compressor, without the need of a conversion specific to the medium. This independence of the media is beneficial as air is the main contaminant, e.g.~entering the system through leaks.

\subsubsection[Helium and compressor cooling]{Cooling of the helium and the compressor}
The heat from the detector will be transferred into the helium. Cooling the helium is performed by a plate heat exchanger, model B12Lx14/1P-SC-S 4x1 1/4" by SWEP\footnote{SWEP International, Landskrona, Sweden, \url{https://www.swep.net/}}.
One side is used for the gas and the second side is used for water as refrigerant. The heat exchanger was over-dimensioned intentionally to obtain a low pressure loss on the gas side. The cooling water is prepared by an HRS040 chiller from SMC\footnote{SMC Corporation, Tokyo, Japan, \url{https://www.smcworld.com/}}, which is water cooled and connected to the PSI cooling water loop. The cooling of the turbo-compressor is also provided by the same chiller.

\subsubsection{Compressor surge protection valve}
The online control system (see below) monitors the current state in the compressor map. If the operating point gets near the surge limit, the pneumatically driven surge protection valve is opened to immediately enforce a low system pressure loss. This prevents the onset of rotating stall or surge. The valve is chosen to be open when inactive for machine safety in case of a lost electrical connection or loss of compressed air. The time constant for a decision is $\leq\qty{1}{\second}$.

\subsubsection{Venturi tube}
\label{sub:Venturi}
\begin{figure*}[h!]
	\centering
	\includegraphics[width=0.7\textwidth]{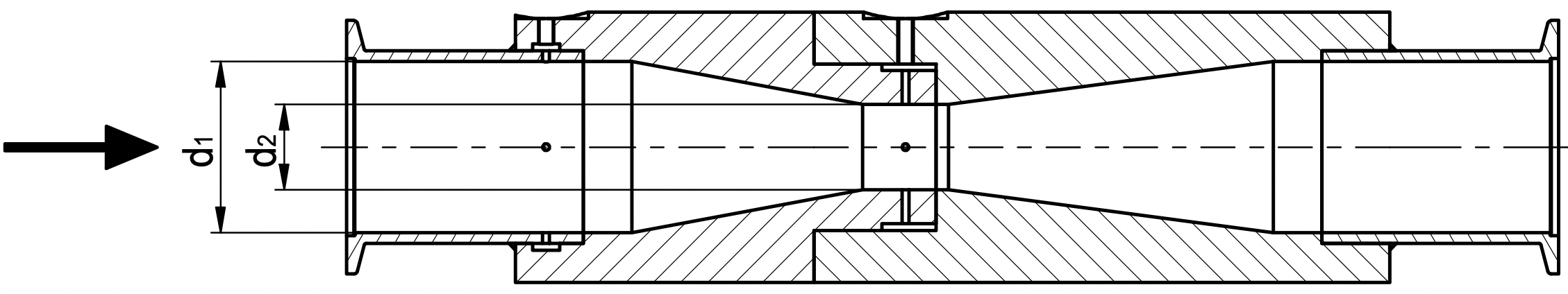}
	\caption{Cross section view of the Venturi tube.
	}
	\label{fig:VT}
\end{figure*}

Many commercially available flow meters are either not optimal for helium or introduce a large pressure loss in the circuit. Venturi tubes in a custom design with carefully selected differential pressure sensors offer an optimum with almost no pressure loss.
The design is shown in \autoref{fig:VT}. The static pressure difference $\Delta p$ between the tube diameter $d_1$ and a constriction in the tube, $d_2$, is a measure for the flow. The mass flow is calculated from $\Delta p$ and the density $\rho$ of the gas (\autoref{eq:Venturi}) using the ratio $\beta =\frac{d_2}{d_1}$ which is $0.5$ in our design.
	\begin{equation}
		\dot{m} =C_D \cdot \varepsilon \cdot \left(\frac{A_2 }{\sqrt{1-\beta^4 }}\right)\cdot \sqrt{2\cdot \Delta p\cdot \rho }
		 \label{eq:Venturi}
	\end{equation}
Here, $C_D$ denotes the flow number (sometimes referred to as $\phi$ in engineering textbooks), $\varepsilon$ being the expansibility factor, $A$ the characteristic area, and $\beta=\frac{d_2}{d_1}$ the diameter ratio of the Venturi tube~\cite{Flucher2020}.

To average out potentially present asymmetric wall pressures, four bores are connected at the circumference with a ring line at both hook-ups for the differential pressure sensor. 
This ring line is integrated in the brass body of the Venturi tube, assembled by brazing.

Calibration of the Venturi flow meters allows to compensate for the manufacturing precision of the Venturi tubes and to correct for any offset or gain errors of the differential pressure sensors.
A \emph{Proline Promass 83A} by Endress+Hauser\footnote{Endress+Hauser, Reinach, Switzerland, \url{https://www.endress.com/}} acts as the master meter for the calibration, which uses the Coriolis method to measure the mass flow directly. The calibration setup is as follows: Compressed helium flows through a pressure reduction valve (to select the operating pressure), then through the master meter, followed by a critical nozzle (creating a choked flow), a straight inlet run, and finally the Venturi tube under test. The setup outlet is hooked up to the PSI helium recovery network to recycle the expensive helium. Using a standard fit procedure, the parameters $C_D$, $ \varepsilon$ and $\frac{A_2 }{\sqrt{1-\beta^4 }}$ are the results of the calibration procedure.

Installed in the cooling setup are two Venturi tubes, before and after the compressor. The one on the input line (\encirc{6} in \autoref{fig:rack_comp}) always sees a flow when the compressor is running independent of the state of the bypass valve. The second one (\encirc{7}) would measure only the remaining flow if the bypass valve is open. With this, both the flow through the compressor and towards the detector system is measured, which are the same under normal operation.

\subsubsection{Sensors}
\paragraph{Temperature and absolute pressure} Test points are located before and after the compressor and after the heat exchanger. The absolute pressure sensors are WIKA type A-10\footnote{WIKA, Klingenberg, Germany, \url{https://www.wika.com/}} in the range \qtyrange{0}{1.6}{\bar}. The temperature sensors are sheated thermocouples (Type K).

\paragraph{Differential pressure} In the beginning, Honeywell TruStability sensors of the HSC series with a measurement range of $\pm\qty{250}{\pascal}$ were used. 
Despite being affordable and easy to use, sudden signal jumps were observed when operated with helium.
We attribute this to the piezoresistive measurement principle and the very low pressure range needed for this application.
Not investigating further, switching sensors resolved the issue. The Setra 267\footnote{Setra Systems, Boxborough, MA, USA, \url{https://www.setra.com/}} and Althen DPS\footnote{ALTHEN GmbH Mess- und Sensortechnik, Kelkheim, Germany, \url{https://www.althensensors.com/}} were evaluated, both in the range \qtyrange{0}{250}{\pascal}. The sensors use either a capacitive (Setra) or an inductive (Althen) measurement principle. During operation, no significant deviation between the readings of the two sensors were observed. The Venturi tubes are connected with stiff silicone tubing to the sensors.

\paragraph{Oxygen concentration} The gas composition is helium with air as a contaminant, mostly from leaks. An Alphasense O2-A2 electrochemical sensor\footnote{Alphasense, Great Notley, Braintree, Essex, United Kingdom, \url{https://www.alphasense.com/}} is installed at the outlet of the experiment, with a \qty{10}{\metre} long return line before the inlet of the compressor unit. This allows to observe any sudden air intake a few seconds before it reaches the compressor, adding valuable time for the control system action. The electrochemical sensor requires amplification. A small sized custom PCB was designed using an INA326 instrumentation amplifier by Texas Instruments in a standard configuration.

\subsubsection{Control system and interlock}
The control system uses the MIDAS slow control system~\cite{MIDAS} with an SCS3000 unit~\cite{SCS3000} for the data acquisition of the sensor data, and to control the compressor, and the bypass valve. All sensors are connected to analogue voltage input terminals in the range of \qtyrange{0}{10}{\volt}, except for the thermocouples which are connected through amplifiers\footnote{Adafruit model 1778 featuring an AD8495 amplifier, \url{https://www.adafruit.com/}}. The compressor is controlled and monitored via analogue signals by the power converter.

A program written in the C language runs on a Raspberry Pi model 4 and supervises the compressor. The pressure ratio $\Pi$ measured across the compressor and the pressure differential $\Delta p$ from the Venturi tube sensors are used to constantly check the state of the compressor and compare with the surge line of the compressor performance map. In case a potentially unsafe state has been reached, the bypass valve is opened immediately. To limit the impeller speed in case of air contamination, the currently maximally allowable speed is calculated from the oxygen sensor reading, assuming a standard oxygen concentration in air of 21\%. For pure air, the speed is limited to one third of the maximal speed in helium. For values in between, linear approximation is in effect.

The Raspberry Pi program generates an interlock signal in case the system does not provide the required helium flow for whatever reason. This signal is used to switch off the electrical power of the detector setup, be it the mock-up or the pixel detector.

\subsection{Test stands}
\label{sec:test_stands}
The helium cooling system as described above is located nearby the setup where the mock-up or detector is installed. In both cases the helium flow is provided by ISO-KF DN50 corrugated stainless steel tubes for short connections up to a few metres long, or suitable PVC tubes for lines up to \qty{10}{\metre}.

\paragraph{Detector mock-up} The mock-up setup is mounted in a frame that can be inserted into an acrylic tube of \mm{230} diameter and \mm{920} length. This provides a sufficiently tight helium chamber, closed with end flanges on both ends. These flanges provide feedthroughs for the helium tubes, the power lines to heat the mock-up detector, and the signal lines of the temperature sensors. 

\paragraph{Pixel detector setup} The detector is mounted in the detector cage as shown in~\cite[Chap.~13]{Mu3e:2020gyw} and operated either inside the Mu3e magnet [ibid., Chap.~4] or in a metal vessel with the same inner shape as the magnet when no field is needed. In this case, the standard feedthroughs of Mu3e for the magnet/vessel are in use.

\section{Results}

\subsection{Compressor performance}
\label{sec:compressor_performance}

The helium plant prototype was used in different configurations, cooling either the \emph{mock-up} or the \emph{\textsc{MuPix10} vertex detector prototype}, which are described in \autoref{sec:test_stands}.

For the silicon heater mock-up, the interface between it and the helium plant is realised with four tubes with \mm{6} inner diameter to feed in the tube connections to the acrylic tube.
The resulting cross section is \qty{113}{\milli\meter\squared} compared to around \qty{200}{\milli\meter\squared} foreseen for the helium ducts in the final experiment.
As a consequence, the differential pressure across the system has a comparatively high value with \qty{146\pm1}{\mbar} for a mass flow of \qty{1.95}{\gps}.
This pressure is already at the upper limit of the compressor and leads to the fact that with this experimental setup no higher mass flows can be measured. 

For the \textsc{MuPix10} vertex detector setup, the interfaces to the Mu3e magnet and the metal vessel housing the detector are realised keeping the cross section at a maximum.
The tubes taper only shortly before the detector to a minimum total cross section of around \qty{200}{\milli\meter\squared}.
Therefore, this setup realistically resembles the final system.
The resulting differential pressure across the system is \qty{36.8\pm0.1}{\mbar} for a mass flow of \qty{1.95}{\gps}.
This enables the operation of the turbo compressor at mass flows beyond \qty{2}{\gps}.

\subsection{Silicon heater detector mock-up}

This mock-up is used for a detailed thermal characterization of the helium cooling system of the Mu3e vertex detector.
With an integrated thermometer on each chip, the temperature within the whole mock-up can be mapped.
The temperature is determined for two specific chip heat loads, which are \mwcm{215} and \mwcm{350}.
The first is the nominal heat dissipation found in \textsc{MuPix10} laboratory studies~\cite{Augustin:2020pkv}.
The second is a conservative limit the cooling system is designed for.

\subsubsection[Measurement procedure]{Measurement procedure and data interpolation} \label{sec:measproc_datainterpolation}

\begin{figure}[h!]
	\centering
	\includegraphics[width=0.5\columnwidth]{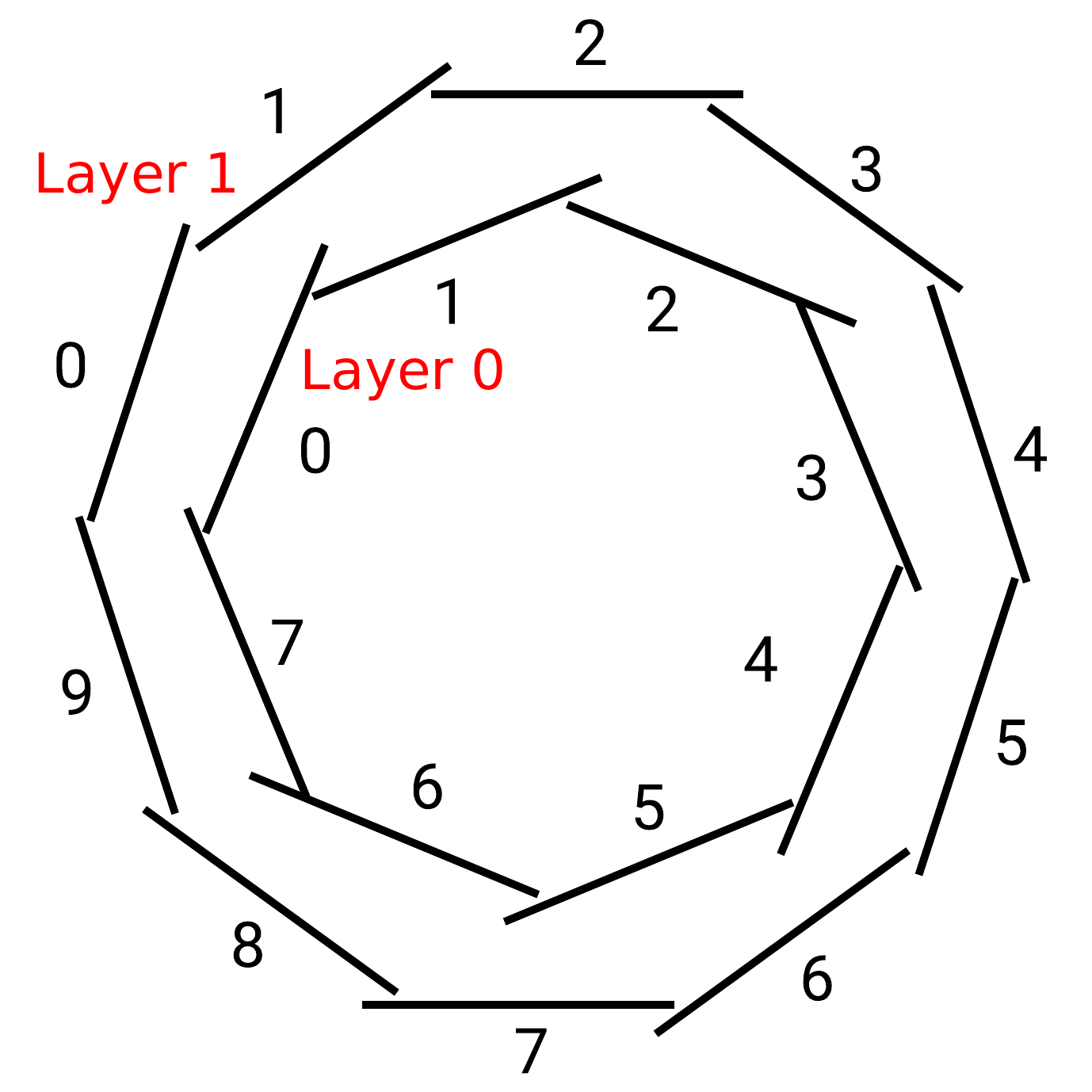}
	\caption{Ladder numbering scheme of the Mu3e vertex detector and the silicon heater mock-up, viewed in the direction of helium flow (upstream to downstream).~\cite{Rudzki:thesis}
	}
	\label{fig:numbering_scheme}
\end{figure}

The measurement cycles are carried out for groups of 3 chips.
Each cycle consists of the following steps: (1) zero-point calibration, (2) providing the heating power in 2 (or more heating) stages for \qtyrange{30}{60}{\second} per heating stage to all heater chips, (3) switching off the heating.
This cycle is repeated until data from all chips has been recorded.
In-between each sequence, power is kept off until the mock-up cools down to the temperature of the gas at the inlet again.
The gas flow remains unchanged at the nominal \qty{1.95}{\gps} during the entire measurement.
The inlet temperature and chip temperature when no power is applied is $T_{ref} = \qty{15.5\pm1.5}{\degreeCelsius}$.
The measured chip temperatures are given as the temperature difference to this reference temperature.
The maximum temperature drift within a measurement window of \qty{120}{\second} is determined to be \qty{0.03}{\kelvin}.
The measurement results are displayed below as temperature maps.
These are 2-dimensional projections of two detector layers. 
The ladder numbering scheme used is displayed in \autoref{fig:numbering_scheme}.

\begin{figure}[t!]
	\centering
	\includegraphics[width=1.05\columnwidth]{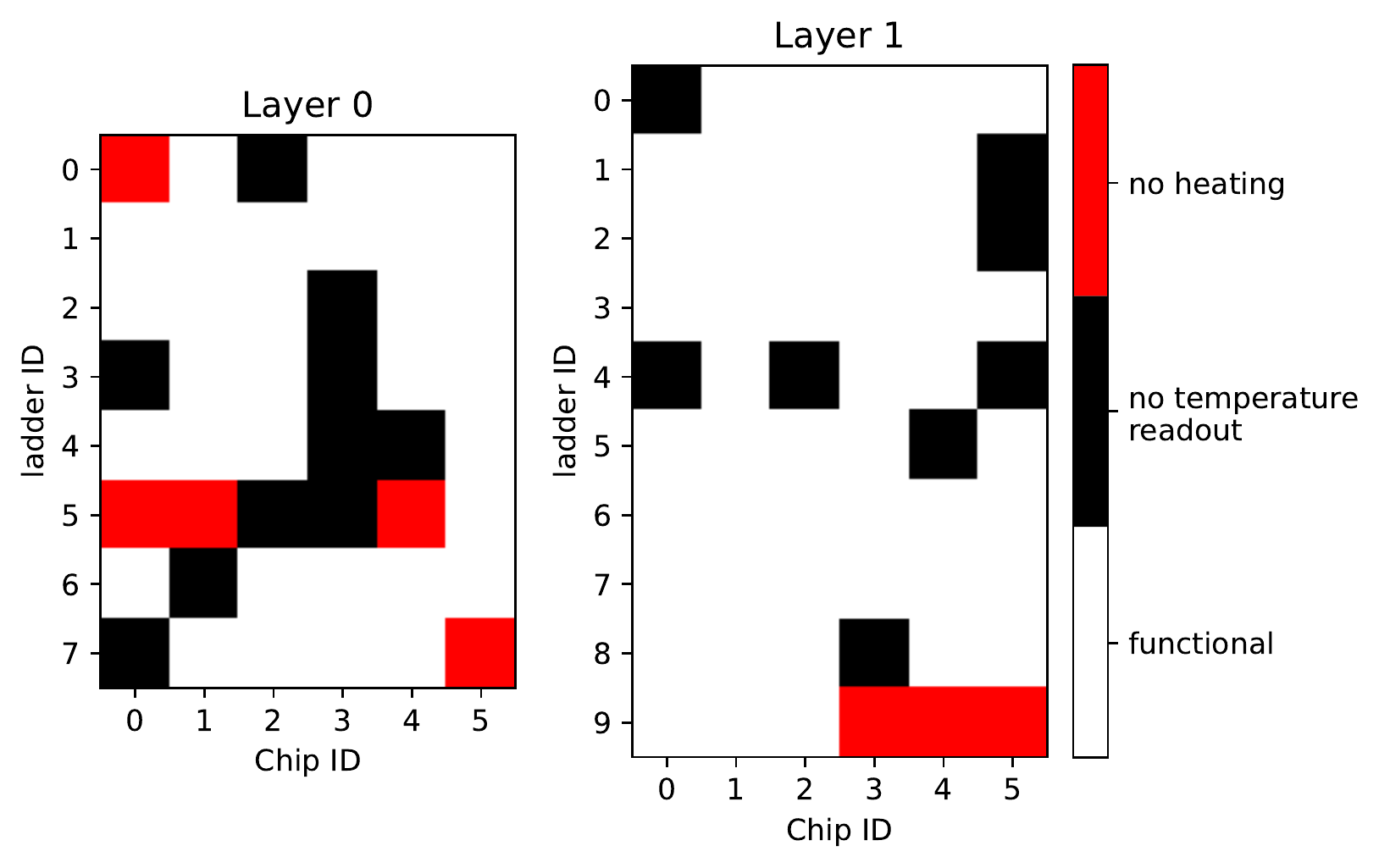}
	\caption{Functionality map for the heater chips of the silicon heater mock-up.
	}
	\label{fig:siHeater_functionalityPattern}
\end{figure}

We could not produce any spare ladders due to the limited availability of HDIs from the supplier.
In consequence, limitations had to be accepted for the ladders used.
No power could be applied to 8 out of 108 chips which results locally in no heating.
The temperature readout is working for 91 of 108 chips.
The pattern of fully functional and limited functional chips is shown in \autoref{fig:siHeater_functionalityPattern}.
The groups of chips with no heating and no functional temperature readout are fully disjoint.

\begin{figure}[t!]
	\centering
	\includegraphics[width=\columnwidth]{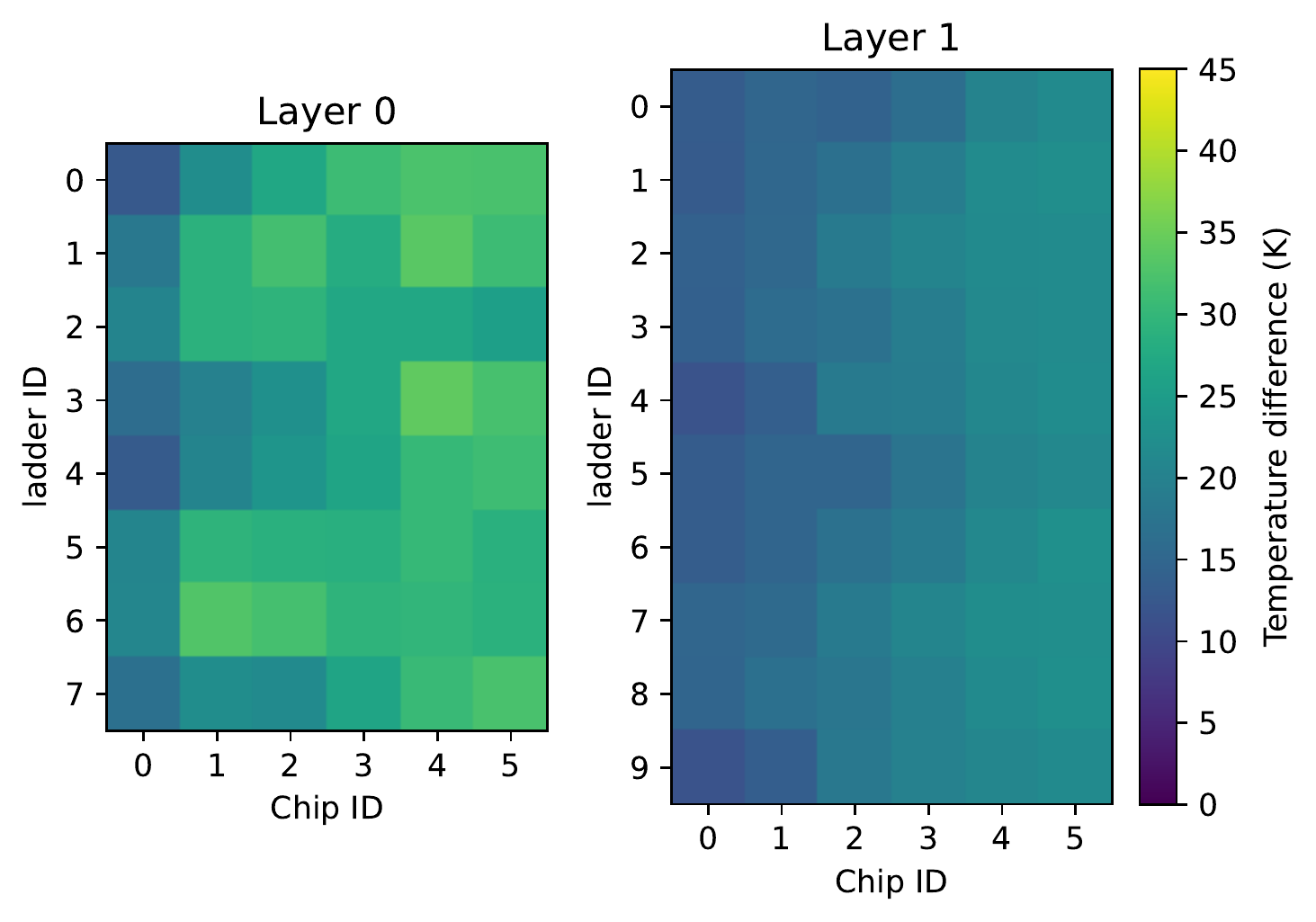}
	\caption{Temperature maps of Layer~0 and 1 for \mwcm{350} uniform heating, obtained by CFD simulation. Temperature difference relative to gas inflow temperature.
	}
	\label{fig:simu_uniform_350mwcm2}
\end{figure}

Interpolation of missing data is based on simulations. 
The same framework from~\cite{Deflorin:thesis} is used to simulate a vertex detector uniformly heated with \mwcm{350}.
The results are shown in~\autoref{fig:simu_uniform_350mwcm2}.

\begin{figure}[h!]
	\centering
	\includegraphics[width=0.485\columnwidth]{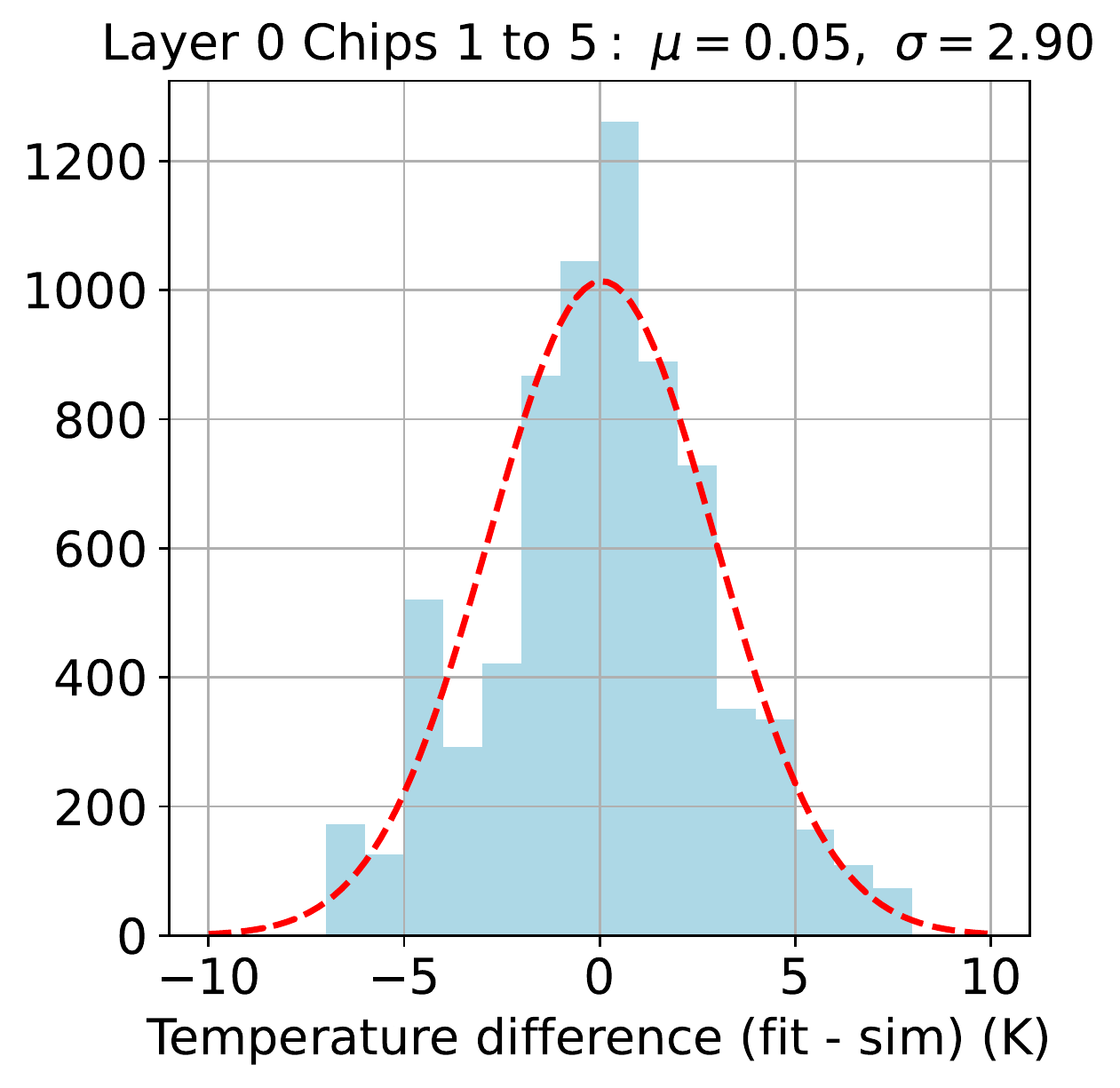}
	\includegraphics[width=0.45\columnwidth]{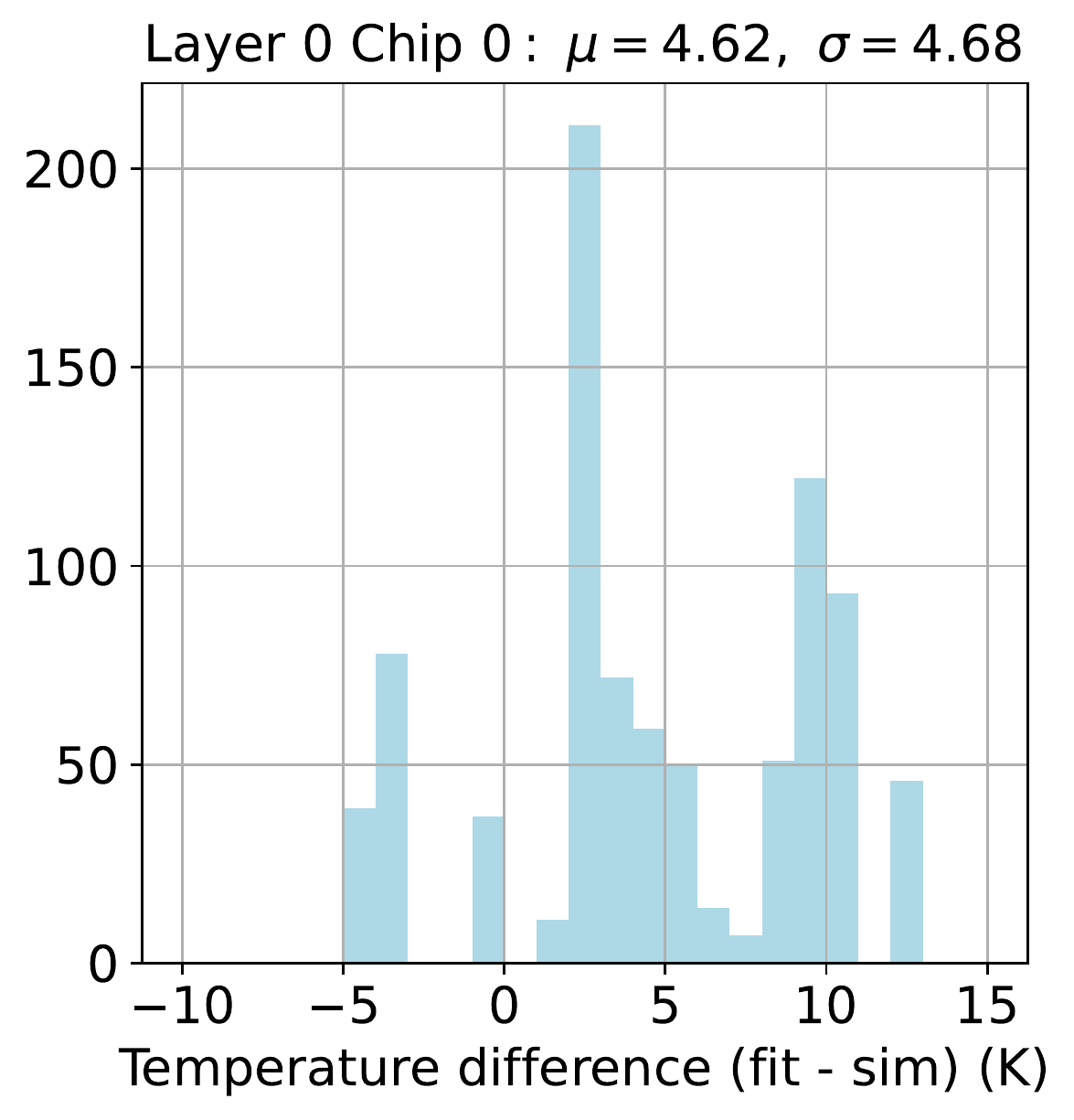}
	\includegraphics[width=0.485\columnwidth]{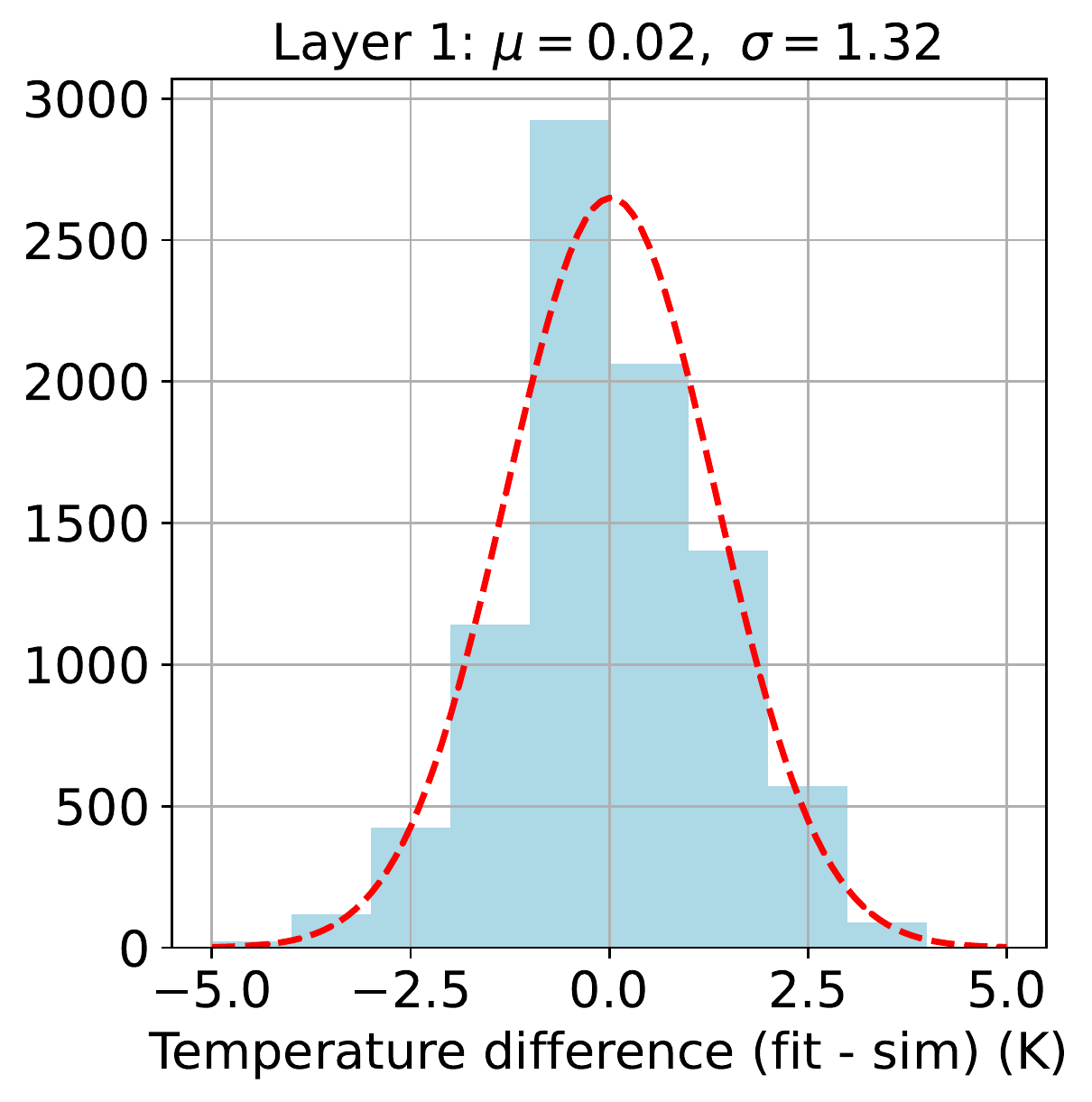}
	\caption{Difference of fitted to simulated chip temperatures for Layer~0 and Layer~1 based on blinding random 17 chips from \autoref{fig:simu_uniform_350mwcm2} and estimating the chip temperatures by a 2nd grade polynomial. 
	Layer~0 is split since the distribution features a left-right asymmetry.
	}
	\label{fig:fit_quality_simu}
\end{figure}

In a Monte-Carlo simulation, 1000 cases of randomised patterns of dysfunctional sensors are generated, with similar patterns to what is observed in the mock-up.
All missing temperature values are then estimated by 2nd grade polynomial fits along the corresponding ladder.
Specific chip patterns are excluded to match the selection criteria of ladders used to construct the mock-up.
These are: (1)~maximum 3 chips without temperature readout per ladder; (2)~maximum 2 neighbouring chips without temperature readout; (3)~at least one chip with chipID~0 or 1 has a functional temperature readout on every ladder, same holds for chipIDs~4 and 5; (4)~in Layer~0 on each ladder, two out of three chips with chipIDs~0 to 2 have a functional temperature readout; (5)~in Layer~0 chipID~5 has always a functional temperature readout.
With this, the accuracy of the fitting method was quantified.
\autoref{fig:fit_quality_simu} shows the histograms of the difference of fitted and initially simulated temperatures.
For Layer~0, the fitted chip temperatures for ChipID~0 have a tendency to be overestimated.
For the remaining chips of Layer~0 and the entire Layer~1, the temperature difference between fit and simulation results in a Gaussian distribution.
With this method, missing temperature values in the simulation can be estimated with an accuracy of \qty{\pm2.9}{\kelvin} for chips~1 to~5 of Layer~0 and \qty{\pm1.3}{\kelvin} for Layer~1. 
The temperatures of chips~0 of Layer~0 are overestimated by this fitting method.
The temperature values of the two chips affected are extracted from the fit, subtracted by \qty{4.6}{\kelvin}, and assigned with an uncertainty of \qty{\pm4.7}{\kelvin}.

The chip thermometers are read out by an SCS3000 unit equipped with Pt1000-measurement cards.
Because of the different temperature coefficients of platinum (assumed by the SCS3000's internal circuitry) and aluminium (used on the silicon heater chips), the temperature output needs translation.
Due to sensor variations, the reference resistance $R_0$ of each thermometer has to be calibrated via

\begin{equation}
	R_0 = \cfrac{\qty{1}{\kilo\ohm} \left( 1 + A T_{SCS} + B T_{SCS}^2\right)}{1 + \alpha T_{ref}}
\end{equation}
\noindent
where $T_{SCS}$ is the temperature output of the SCS3000 assuming a Pt1000 input, $A = \qty{3.9083E-3}{\degreeCelsius^{-2}}$ and $B = \qty{-5.775E-7}{\degreeCelsius^{-2}}$ are the temperature coefficients of platinum~\cite{Pt1000}, $\alpha = \SI{4.17\pm0.02E-3}{\degreeCelsius^{-1}}$ is the temperature coefficient of the resistive thermometers made from aluminium~\cite{Tormann:thesis}, and $T_{ref} = \qty{15.5\pm1.5}{\degreeCelsius}$ is the reference chip temperature when no power is applied.

When heating the mockup, the actual measured temperatures $T_{chip}$ are then determined by 

\begin{equation}
	T_{chip} = \cfrac{1}{\alpha} \left( \cfrac{\qty{1}{\kilo\ohm} \left( 1 + A T_{SCS} + B T_{SCS}^2\right)}{R_0} -1 \right).
\end{equation}

Thus, the uncertainty of the temperature $T_{chip}$ is given by the coefficient $\alpha$ and the reference temperature $T_{ref}$. 

\begin{figure}[h!]
	\centering
	\includegraphics[width=\columnwidth]{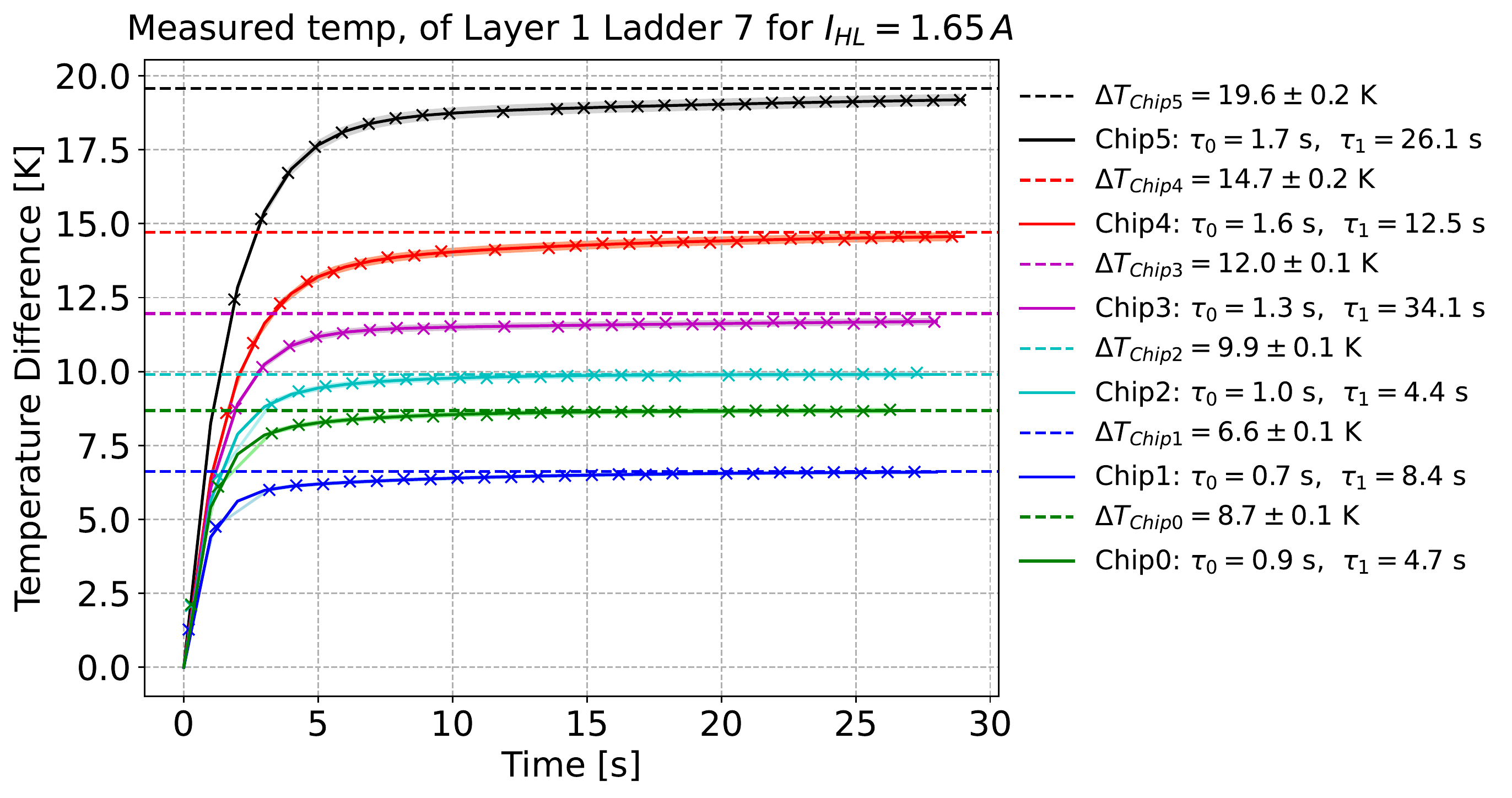}
	\caption{Transient temperature measurement for one ladder of Layer~1 of the silicon heater mock-up. Both half-ladders are powered by providing \qty{1.65}{\ampere}. The crosses show the measured temperature for every chip. The solid line shows the double-exponential fit with the temperature uncertainty (which results from the thermometer gauge and the temperature coefficient $\alpha$) indicated as light band around it. The dashed line shows the equilibrium temperature obtained from the fit (based on~\cite{Rudzki:thesis}, reprocessed data and added error band).
	}
	\label{fig:transient_l1-7_raw}
\end{figure}

The final temperature measurements for every chip are obtained by the transient temperature curves (e.g.~\autoref{fig:transient_l1-7_raw}) when applying the power to the mock-up.
As discussed in~\cite{Rudzki:thesis}, the warm up process can be described by the sum of two exponentials:
\begin{equation}
	T(t) = T_a \left(1-e^{-(t-t_0)/\tau_0}\right) + T_b \left(1-e^{-(t-t_0)/\tau_1}\right)
\end{equation}
with $T_a$ and $T_b$ the temperature contributions of the two time constant $\tau_0$ and $\tau_1$, and $t_0$ the time offset accounting for the start-up time.
The fast time constant results from the self-heating of the light pixel ladders, while the slow time constant comes from the heat-up of the support and supply structure, which are more massive.
The sum $T_{chip} = T_a +T_b$ is then considered as the equilibrium chip temperature for the heat load applied.

\subsubsection{Temperature maps}
\label{sec:tmaps}

\begin{figure}[h!]
	\centering
	\includegraphics[width=0.8\columnwidth]{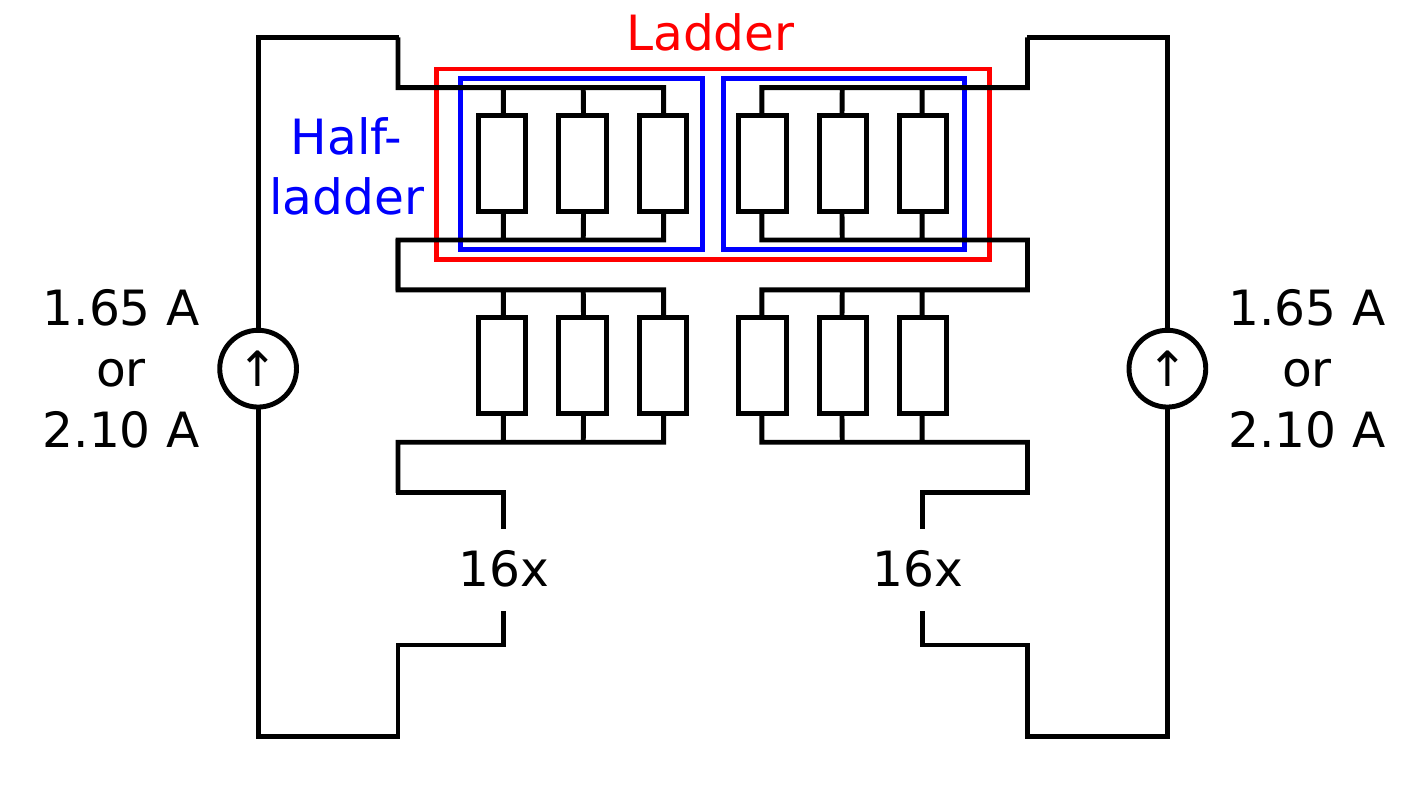}
	\caption{Sketch of the powering scheme of the silicon heater mock-up. On each side, all 18 half-ladders are powered in series. The chips of a half-ladder are powered in parallel. 
	}
	\label{fig:siHeater_schematic_powering}
\end{figure}

The measurement method described above is applied to all chips with functional readout with two heating stages.
The chips of each half-ladder (chips 0 to 2 or chips 3 to 5 of a ladder) are powered in parallel.
On each side of the mock-up, all such half-ladders are powered in series with a constant current (see~\autoref{fig:siHeater_schematic_powering}).
The currents supplied are \qty{1.65}{\ampere} or \qty{2.1}{\ampere}, respectively.
Under the assumption of a constant temperature coefficient~$\alpha$, this would result in a heat load of \mwcm{182} or \mwcm{295}, respectively, with a half-ladder resistance of $R_{HL} = \qty{2.75}{\ohm}/3 \approx \qty{0.92}{\ohm}$ at room temperature and a heated surface of $3 \times \left(1.99 \times \qty{2.30}{\centi\meter^2} \right)$.

However, as the heating resistance of the heater chips changes with their temperature, the effective heat load is higher.
In addition, the current flow varies from chip to chip due to their different resistances, which are influenced by their respective temperatures.
Furthermore, the resistances on the HDI supply lines to each chip differ.
To obtain chip temperatures assuming a uniformly heated mock-up, two correction factors $f_R$ and $f_I$ for the resistance and current of a chip are introduced:
\begin{equation}
	f_R = \left(1+\alpha \Delta T_{chip}\right) 
\end{equation}
\begin{equation}
	f_I = \cfrac{1}{1+\alpha \Delta T_{chip} + \cfrac{R_{chip,HDI}}{\qty{2.75}{\ohm}}} \times \cfrac{3}{\sum \cfrac{1}{1+\alpha \Delta T_i + \cfrac{R_{i,HDI}}{\qty{2.75}{\ohm}}}}
\end{equation}
with $\Delta T_{chip}$ the temperature difference of the chip between powered to unpowered state, and $R_{i,HDI}$ the different resistances on the supply lines to each chip with the measured values $R_{0\&5,HDI} = \qty{18}{\milli\ohm}$, $R_{1\&4,HDI} = \qty{45}{\milli\ohm}$, and $R_{2\&3,HDI} = \qty{65}{\milli\ohm}$~\cite{Rudzki:thesis}.

Assuming a linear dependence of temperature to heating power, the measurement results can be translated to heat loads $\phi_{0} = \mwcm{215}$ and $\phi_{lim} = \mwcm{350}$, by the factor:
\begin{equation}
	f_P = \cfrac{\phi_i}{\phi_{nom}}
\end{equation}
The suitability of the assumption that temperature and heat dissipation scale linearly is discussed in section~\ref{sec:TtoP}

\begin{figure}[t!]
	\centering
	\includegraphics[width=\columnwidth]{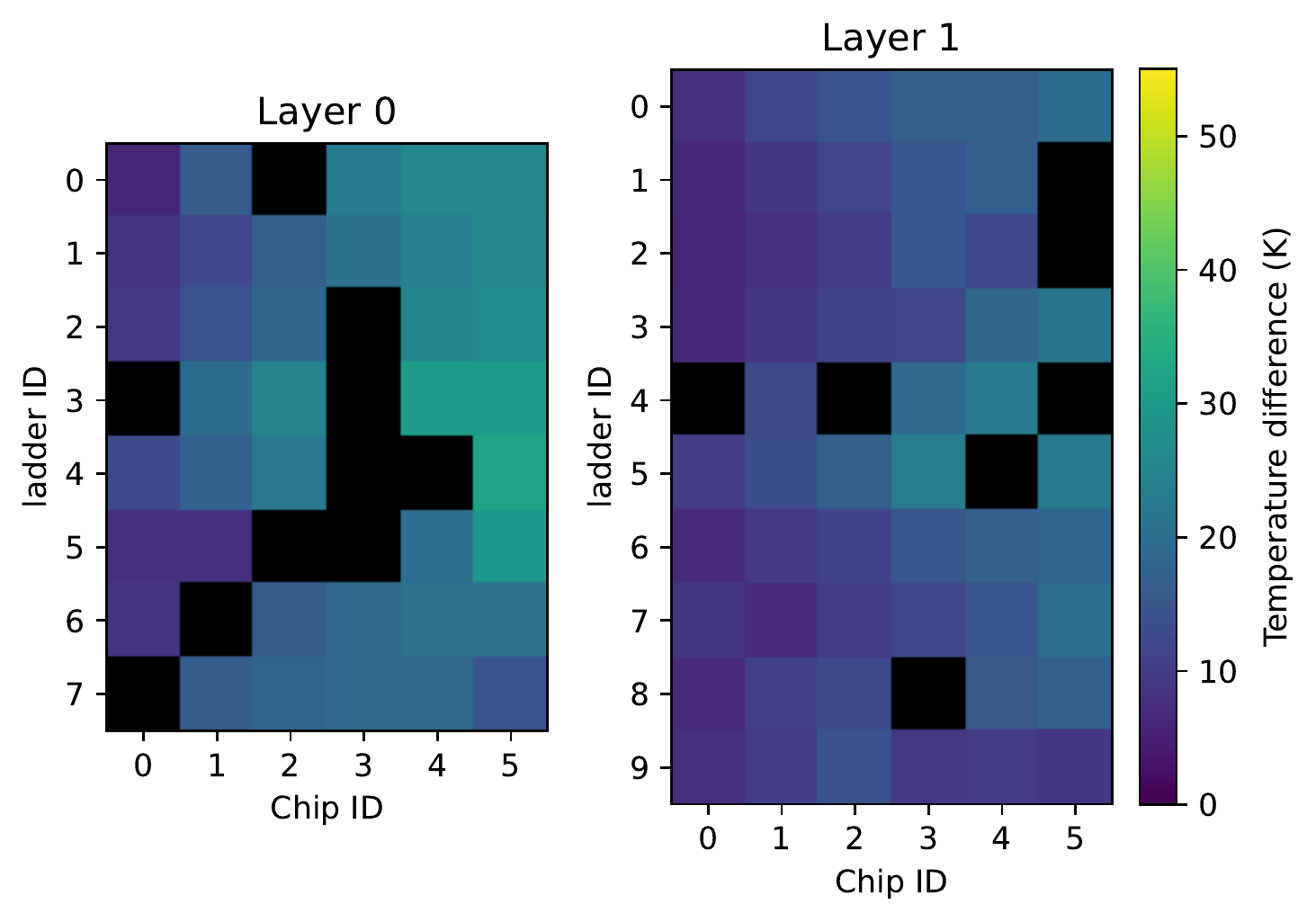}
	\caption{Measured raw temperature maps of Layer~0 and 1 for powering all half-ladders in series with a heating current of \qty{1.65}{\ampere}.
	Temperature difference relative to gas inflow temperature.
	}
	\label{fig:siHeater_tMap_raw_1650mA}
\end{figure}

\begin{figure}[t!]
	\centering
	\includegraphics[width=\columnwidth]{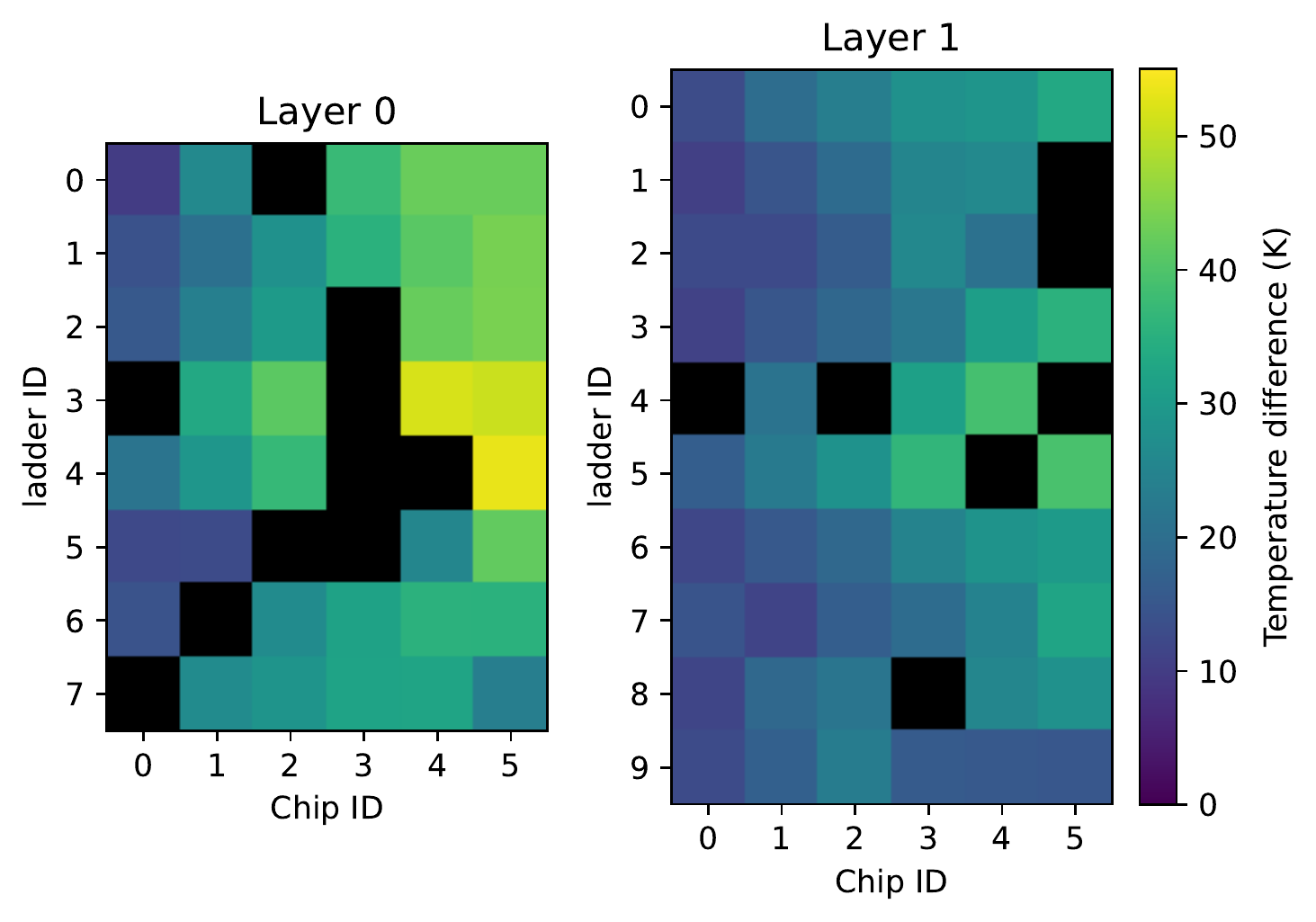}
	\caption{Measured raw temperature maps of Layer~0 and 1 for powering all half-ladders in series with a heating current of \qty{2.10}{\ampere}. 
	Temperature difference relative to gas inflow temperature.
	}
	\label{fig:siHeater_tMap_raw_2100mA}
\end{figure}

\begin{figure}[h!]
	\centering
	\includegraphics[width=\columnwidth]{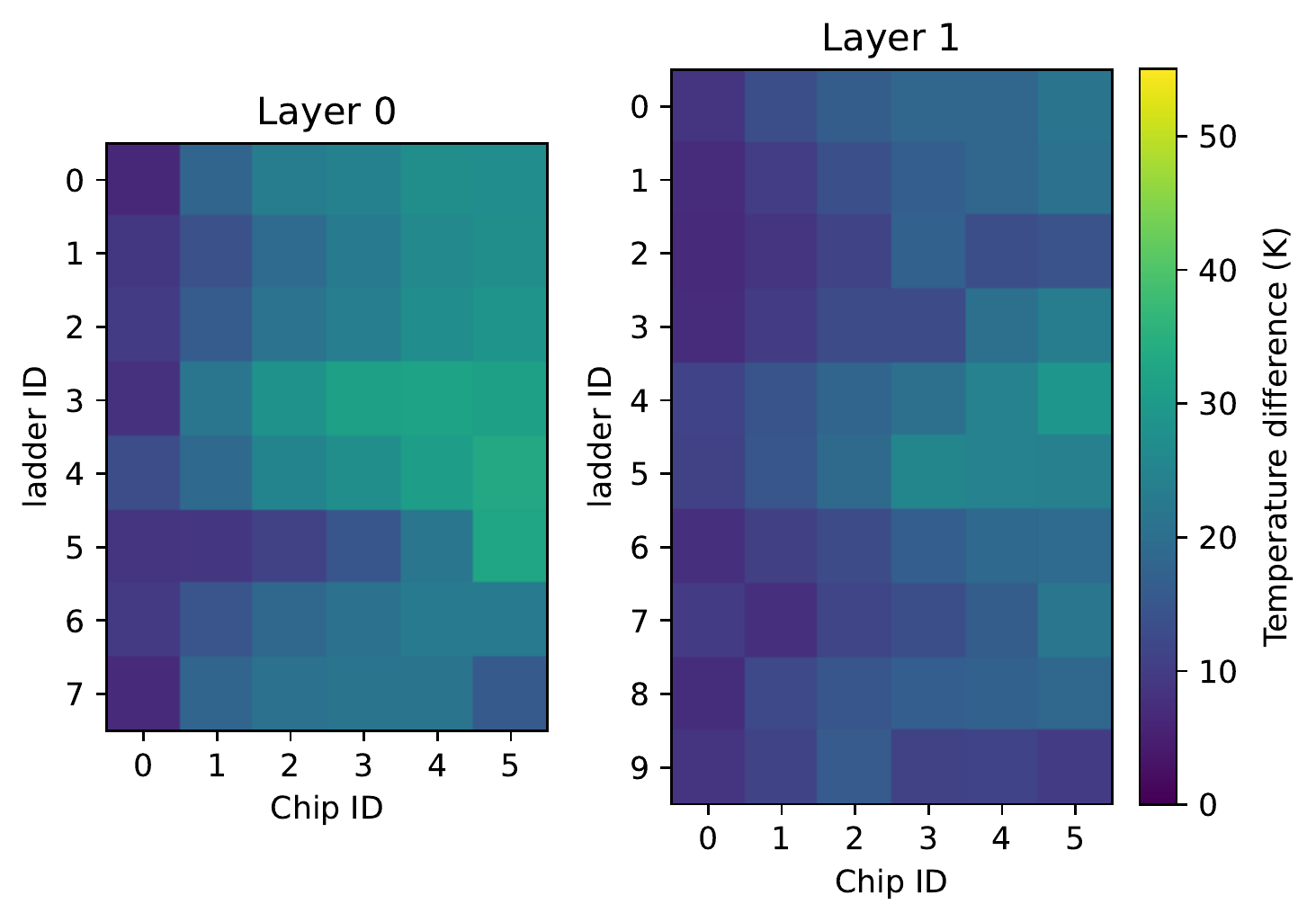}
	\caption{Temperature maps of Layer~0 and 1 for a heat load of \mwcm{215}. Based on \autoref{fig:siHeater_tMap_raw_1650mA}. Missing chip temperatures estimated by 2nd grade polynomial fits along the corresponding ladder. Correction factors $f_R^{-1}$, $f_I^{-2}$ and $f_P$ are applied to determine temperature for a uniformly heated mock-up.
	}
	\label{fig:siHeater_tMap_215mWcm2}
\end{figure}

\begin{figure}[h!]
	\centering
	\includegraphics[width=\columnwidth]{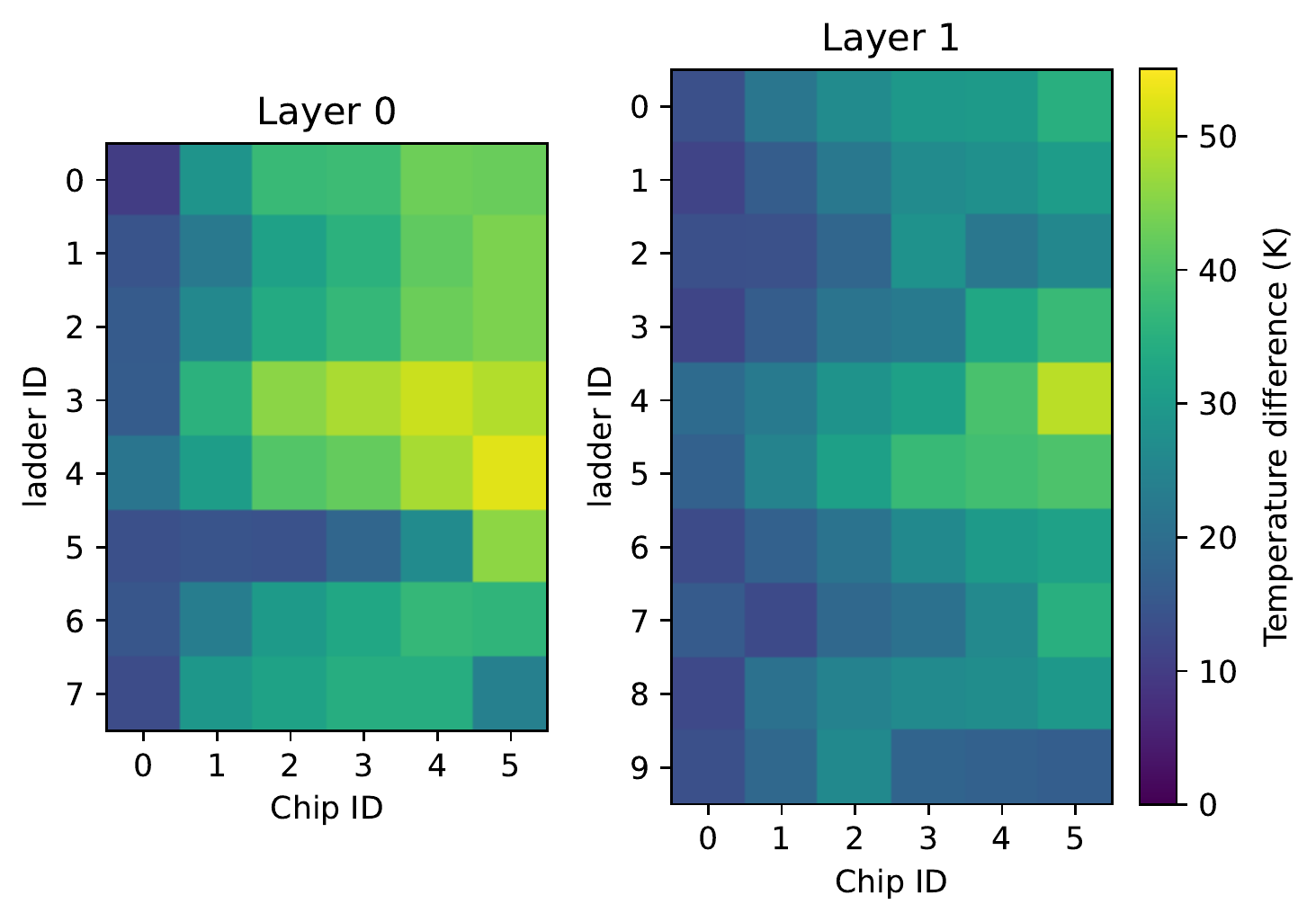}
	\caption{Temperature maps of Layer~0 and 1 for a heat load of \mwcm{350}. Based on \autoref{fig:siHeater_tMap_raw_2100mA}. Missing chip temperatures estimated by 2nd grade polynomial fits along the corresponding ladder. Correction factors $f_R^{-1}$, $f_I^{-2}$ and $f_P$ are applied to determine temperature for a uniformly heated mock-up.
	}
	\label{fig:siHeater_tMap_350mWcm2}
\end{figure}

Starting from the raw data shown in Fig.~\ref{fig:siHeater_tMap_raw_1650mA} and~\ref{fig:siHeater_tMap_raw_2100mA}, the following analysis steps are applied:
(1) estimation of missing chip temperature by a 2nd grade polynomial fit along the corresponding ladder (described in \autoref{sec:measproc_datainterpolation}), (2) application of the correction factors $f_R^{-1}$, $f_I^{-2}$ and $f_P$ for each chip.
This way, temperature maps for a uniformly heated mock-up are obtained for the heat loads $\phi_0$ and $\phi_1$ which are shown in Fig.~\ref{fig:siHeater_tMap_215mWcm2} and~\ref{fig:siHeater_tMap_350mWcm2}. 
A prominent hot region is visible in both Layers~0 and~1.
This feature is expected from a vortex forming inside the gas channels as predicted in~\cite{Tormann:thesis} and confirmed in~\cite{Deflorin:thesis}.

The average temperature and the maximum temperature reached are summarised in \autoref{tab:siHeater_temp}.
All chip temperatures are well below the glass transition temperature of the adhesives used, which is $T_{crit} = \qty{70}{\degreeCelsius}$ (assuming an inlet temperature of \qty{0}{\degreeCelsius}).


\begin{table}[h]
	\centering
	\begin{tabular}{llll}
		\toprule
		& Heat load & Layer 0 & Layer 1      \\
		\midrule
		\multirow{2}{*}{$T_{mean}$} & \mwcm{215} & \qty{20.3\pm0.7}{\kelvin} & \qty{14.8\pm0.2}{\kelvin} \\
		& \mwcm{350} & \qty{32.0\pm0.7}{\kelvin} & \qty{24.1\pm0.3}{\kelvin} \\
		\midrule
		\multirow{2}{*}{$T_{max}$} & \mwcm{215} & \qty{33.2\pm1.2}{\kelvin} & \qty{28.9\pm1.5}{\kelvin} \\
		& \mwcm{350} & \qty{52.5\pm0.9}{\kelvin} & \qty{49.3\pm1.5}{\kelvin} \\
		\bottomrule
	\end{tabular}
	\caption{Average and maximum chip temperatures of the two tracking layers for heat loads of \mwcm{215} and \mwcm{350}, based on \autoref{fig:siHeater_tMap_215mWcm2} and~\ref{fig:siHeater_tMap_350mWcm2}.
	The errors include the fit uncertainties for estimating missing chip temperatures, as well as systematic uncertainties in the temperature measurement and the correction factors $f_R^{-1}$ and $f_I^{-2}$ originating from $\alpha$ and $T_{ref}$.} 
	\label{tab:siHeater_temp}
\end{table}


\subsubsection{Temperature-to-power relation}
\label{sec:TtoP}

\begin{figure}[h!]
	\centering
	\includegraphics[width=\columnwidth]{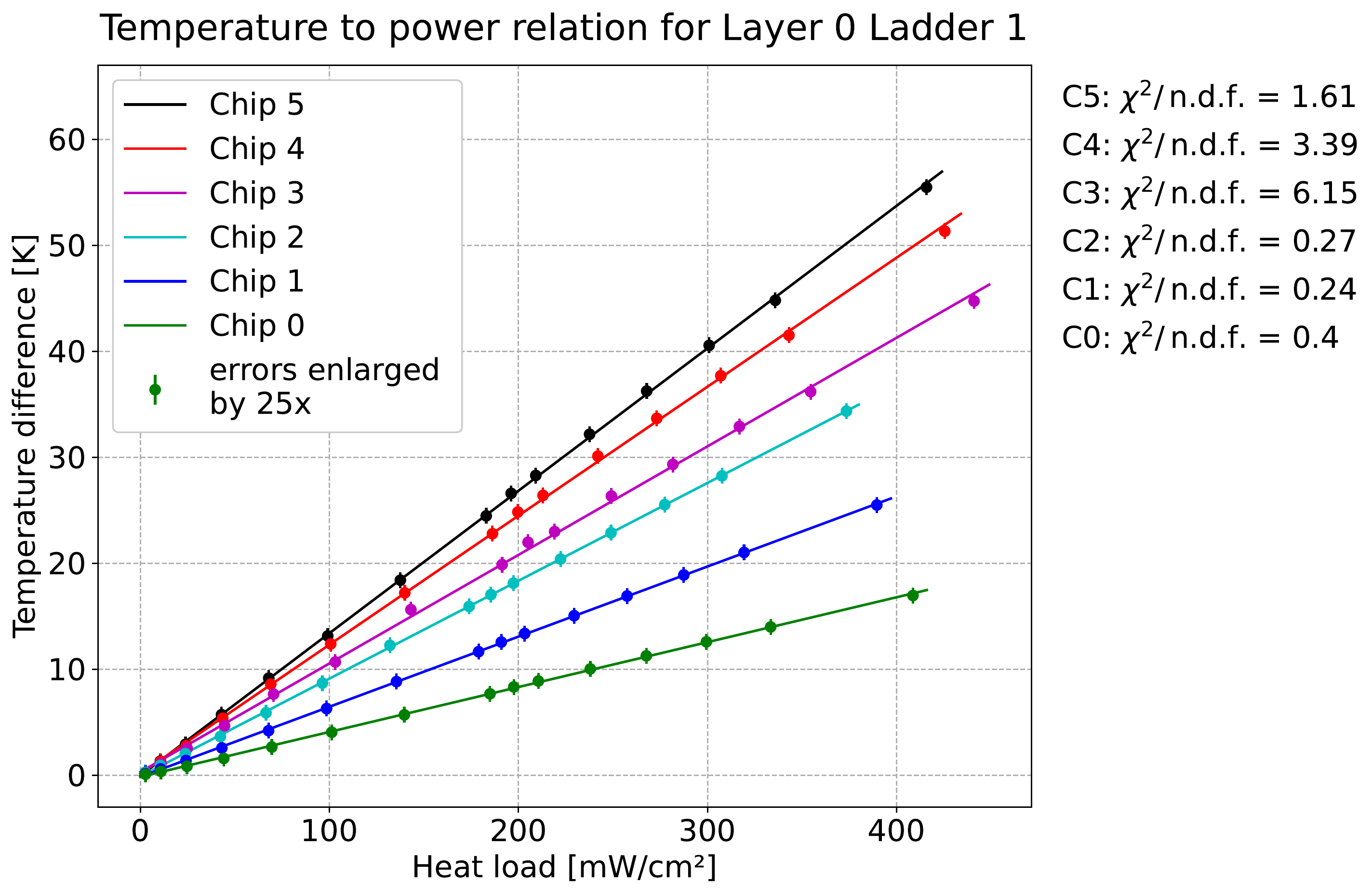}
	\includegraphics[width=\columnwidth]{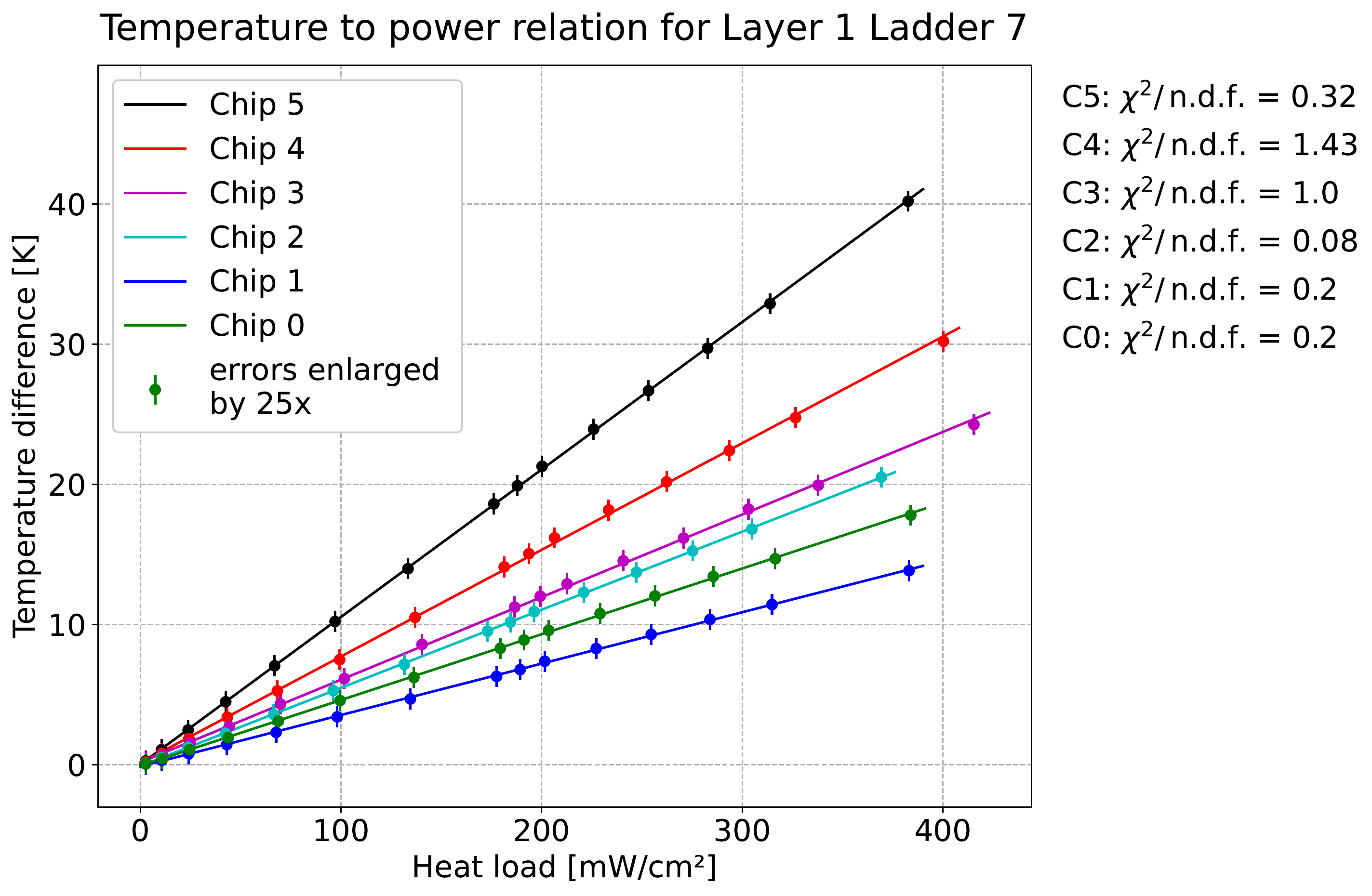}
	\caption{Temperature-to-power relations for Ladder~1 of Layer~0 and Ladder~7 of Layer~1 in the silicon heater mock-up. The measured raw temperature is plotted against the corrected heat load which accounts for thermal resistance changes and different currents provided to the individual heater chips.
	The uncertainty on the temperature is the maximum expected temperature shift of the system during the measurement. 
	}
	\label{fig:siHeater_linearity}
\end{figure}

In the previous section, the assumption of a linear dependence of temperature to heating power was made to apply the correction factor $f_P$.
The relation is investigated by an extended series of measurement cycles with 13 additional power values for one ladder in Layer~0 and Layer~1 each.
The measurement results are shown in \autoref{fig:siHeater_linearity}. 
In this measurement, the dependency of temperature to heating power is of interest.
Thus, the measured raw temperature is plotted against the effective heat load applied.
As a result, the correction factors $f_R$ and $f_I^2$ introduced above are applied to the heat dissipation.
For each heating stage, the heat dissipation on the individual chip varies due to these corrections.



For all chips of the ladders studied, the temperature-to-power relation is found to be linear.
This confirms the previously made assumption that a measured temperature can be translated to a different heat load by the correction factor $f_P$.

\subsubsection[Temperature vs. mass flow]{Temperature as function of the mass flow}

\begin{figure}[h!]
	\centering
	\includegraphics[width=\columnwidth]{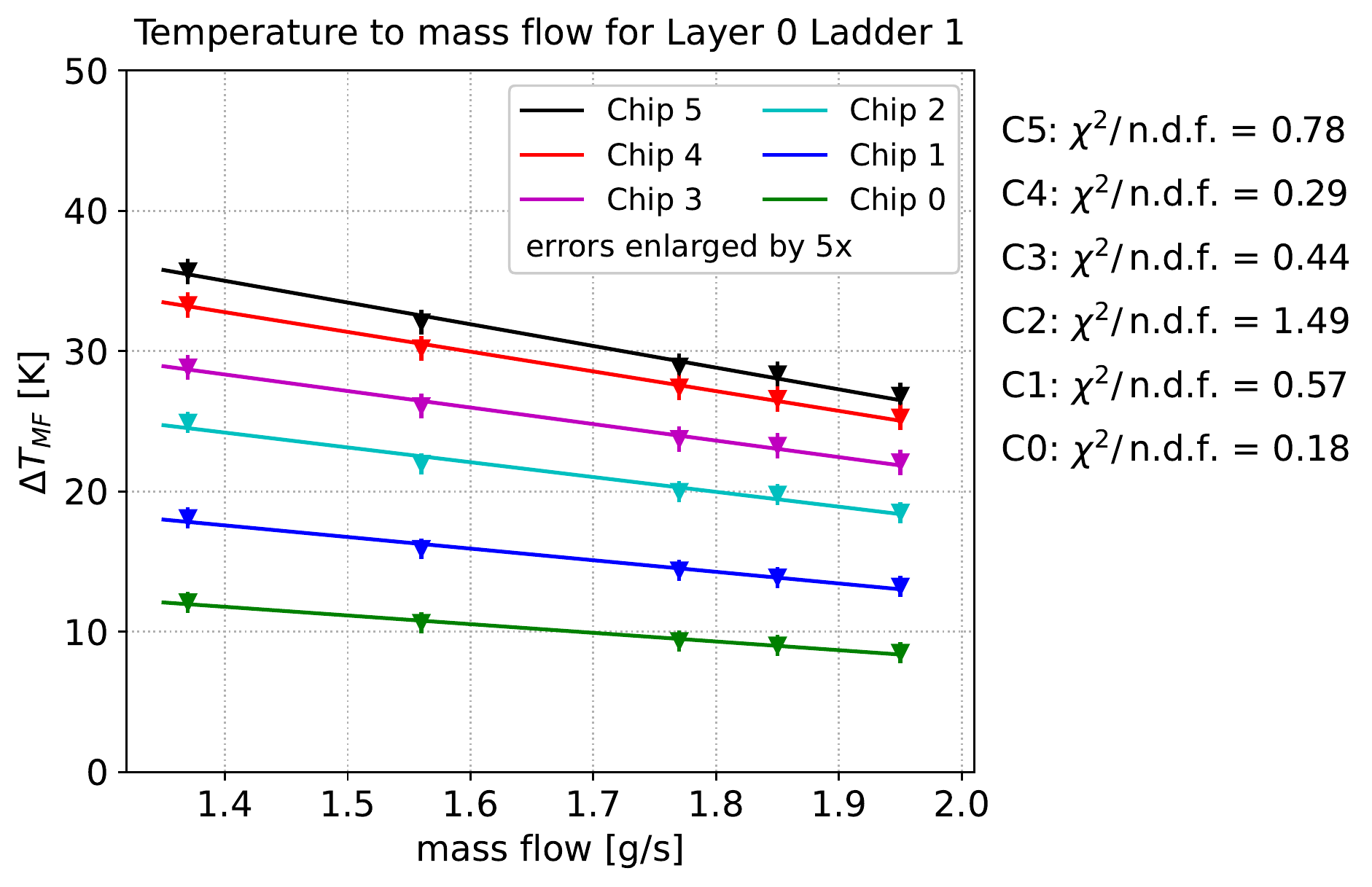}
	\includegraphics[width=\columnwidth]{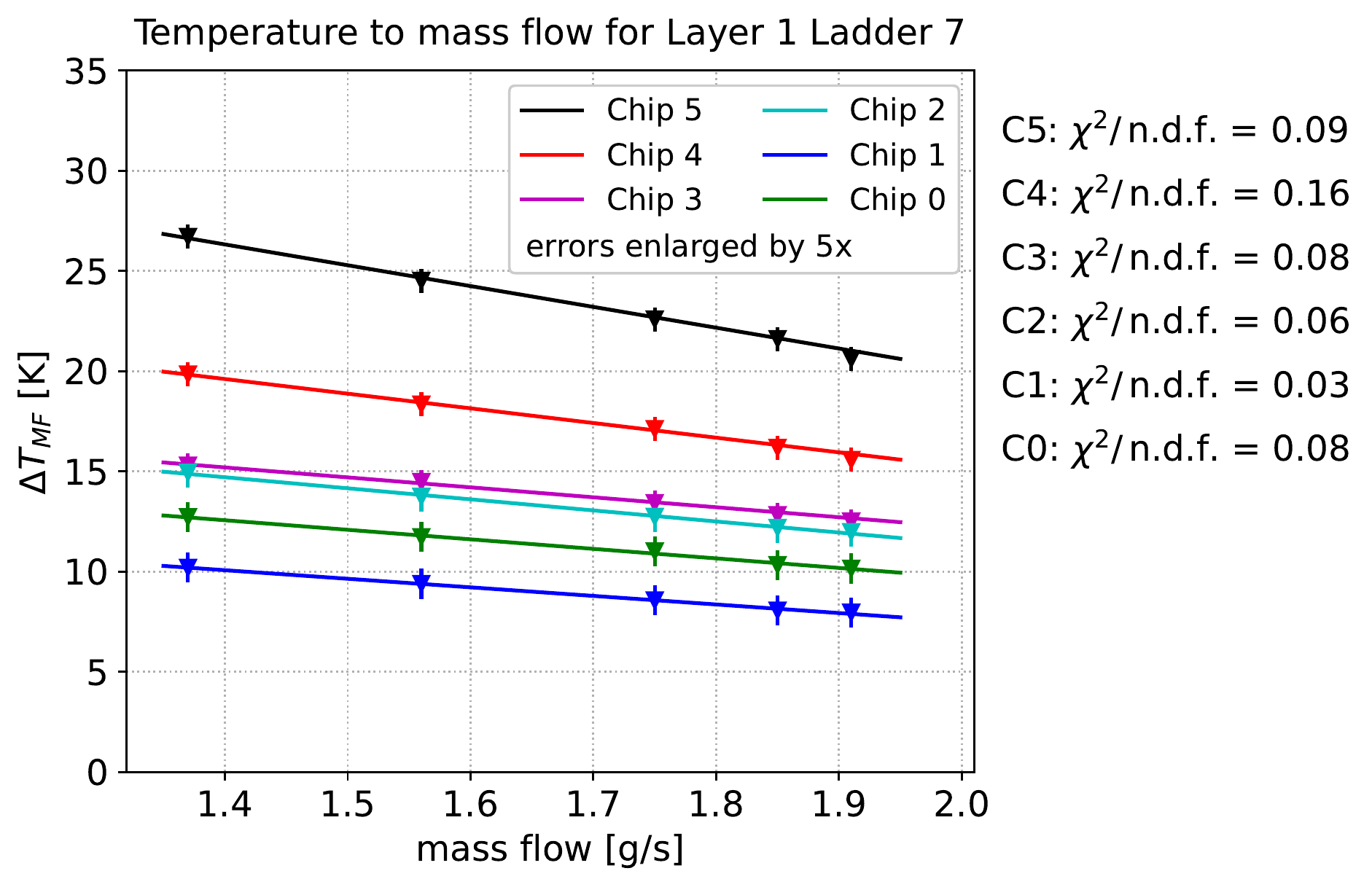}
	\caption{Temperature-to-mass flow relations for Ladder~1 of Layer~0 and Ladder~7 of Layer~1 in the silicon heater mock-up. 
	Correction factors $f_R^{-1}$, $f_I^{-2}$ and $f_P$ are applied to determine temperature for a uniformly heated mock-up.
	The uncertainty on the temperature is the maximum expected temperature shift of the system during the measurement.
	A linear fit is applied to every chip.
	}
	\label{fig:siHeater_TtoMF}
\end{figure}

The operating range of the cooling system is studied for mass flows below the nominal mass flow around \qty{2}{\gps}.
As discussed in \autoref{sec:compressor_performance}, the differential pressure on the helium line was at the limit to operate the turbo compressor preventing mass flow above \qty{2}{\gps}.
The chip temperatures are measured for different mass flows provided by the helium plant for one ladder of both layers each.
The results are shown in \autoref{fig:siHeater_TtoMF}.

A linear fit is applied to the temperature of every chip.
For all chips in both layers, the chip temperature depends linearly on the mass flow for the range of \qtyrange{1.35}{1.95}{\gps}.

The expected chip temperatures in the vertex detector can thus be approximated for the case of reduced compressor speeds.
The linear dependency for higher mass flows than \qty{2}{\gps} has to be confirmed with the final system, which allows a lower differential pressure in the system. 
A study with higher mass flows has been performed with the vertex detector prototype, which however has a different geometry and flow distribution. 
These results are discussed in the next section.

\subsection{MuPix10 vertex detector prototype} \label{sec:VtxPrototype}

The vertex detector prototype has been successfully operated in a helium atmosphere inside the Mu3e magnet for several days.
The power dissipation for the \textsc{MuPix10} chips has been ranging from \qtyrange{210}{225}{\milli\watt/\centi\meter^2} with smaller deviations between the individual chips.

While \textsc{MuPix10} has an integrated temperature diode, it could not be used for a temperature measurement.
The position of the analogue diode is too close to a heat source in the chip periphery.
Two additional temperature sensitive circuits, which are designed to be read out via the data stream, failed due to a bug in the Mu3e slow control interface of this chip.
The problem is resolved on the final chip version, the \textsc{MuPix11} (which was not available at the time of this study).

\begin{figure}[t!]
	\centering
	\includegraphics[width=0.7\columnwidth]{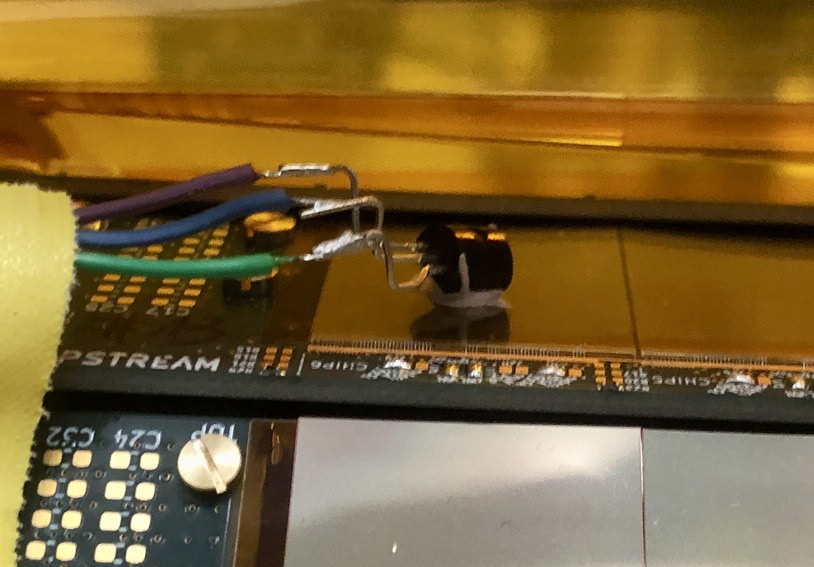}
	\caption{LM35 temperature sensor glued on the active pixel matrix of Chip~0 of Ladder~2 in Layer~1 of the vertex detector prototype.
	}
	\label{fig:lm35}
\end{figure}
\begin{figure}[t!]
	\centering
	\includegraphics[width=0.7\columnwidth]{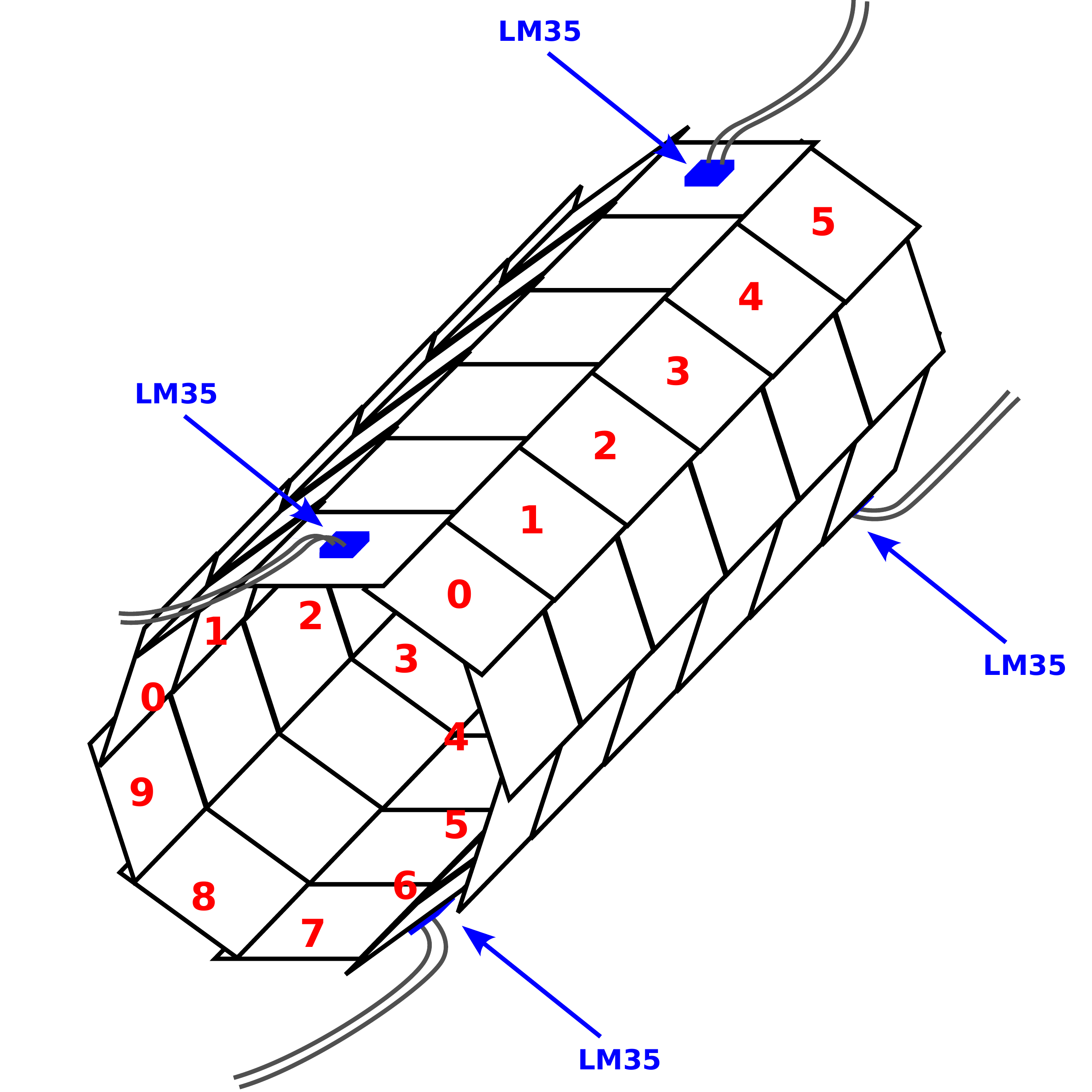}
	\caption{Positions of the four installed LM35 temperature sensors in the vertex detector prototype. All sensors were installed in the outer Layer~1.
	}
	\label{fig:lm35_position}
\end{figure}

Instead, for a temperature measurement of the vertex detector prototype, LM35 temperature sensors\footnote{Texas Instruments, LM35 precision centigrade temperature sensors, \url{https://www.ti.com/lit/ds/symlink/lm35.pdf}} are glued on the active pixel matrix of four selected chips with thermally conductive glue (\autoref{fig:lm35}).
The locations are on Ladders~2 and~6 of Layer~1, both on chips~0 and~5, as visualised in~\autoref{fig:lm35_position}.

\begin{figure}[t!]
	\centering
	\includegraphics[width=0.7\columnwidth]{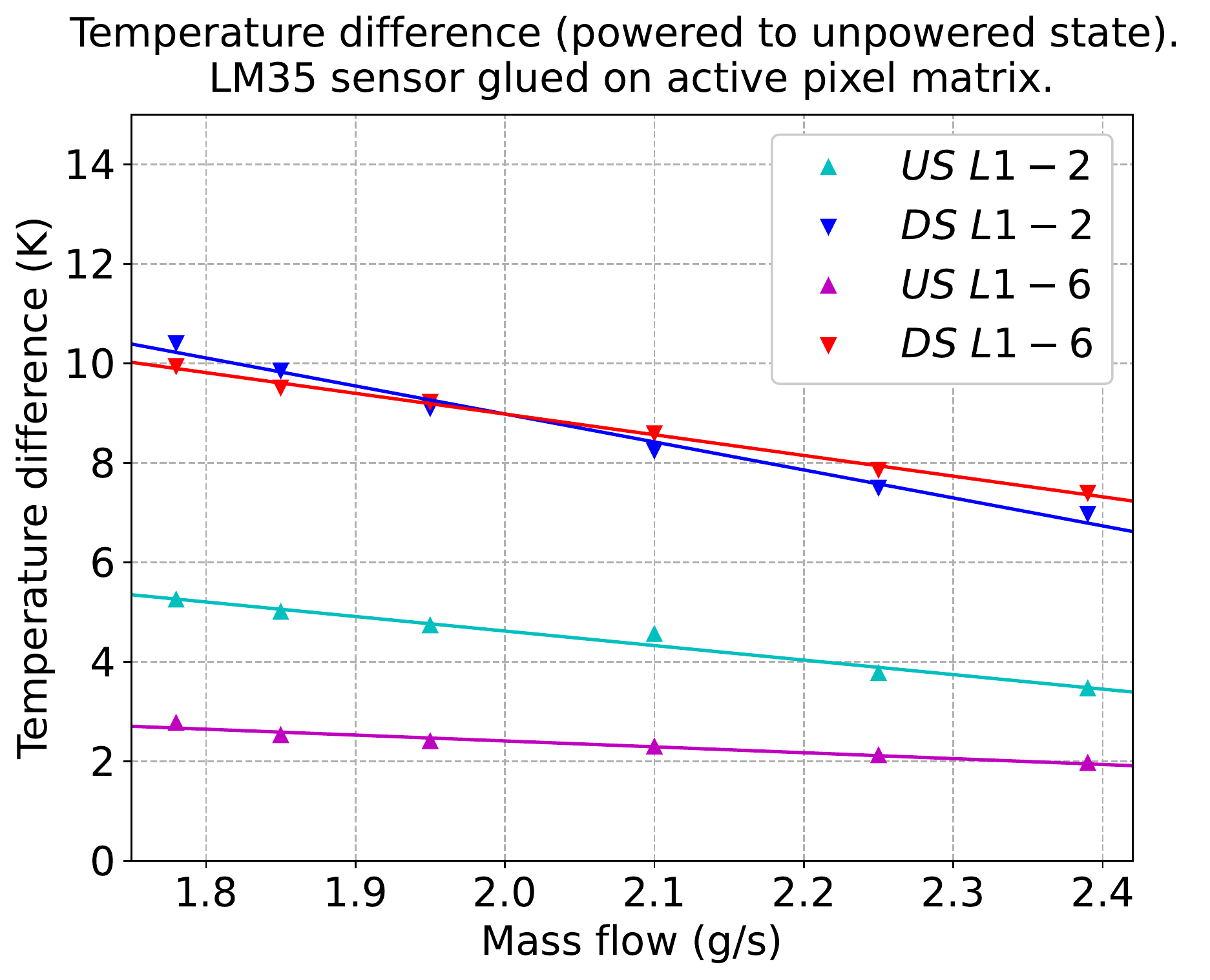}
	\caption{Temperature measured by LM35 sensors glued onto the active matrix of 4 chips in the \textsc{MuPix10} vertex detector prototype for different helium mass flows. 
	}
	\label{fig:MP10_prototype_temperatures}
\end{figure}

The temperatures are measured for different mass flows and are given as differences to the inlet gas temperature.
The results are shown in \autoref{fig:MP10_prototype_temperatures}.

As the modified geometry results in a different helium distribution and the PCB-based ladders differ in their property as a heat sink for the chip periphery, the absolute temperature values can not be compared directly to the results from the silicon heater mock-up.
As with the silicon heater mock-up, the temperature scales linearly with the helium mass flow in this setup.
Due to the optimised cross section of the helium ducts, mass flows up to \qty{2.4}{\gps} could be covered.

\paragraph{DAQ integration and cosmic runs}
The vertex detector prototype was operated in two dedicated measurement campaigns.
The main goal was to prove that the readout is able to handle the high data rate produced by the vertex detector, which is a key for the Mu3e experiment.
In the DAQ integration run 2021~\cite{Rudzki:thesis, Koppel:2022kbd}, correlation data between the two tracking layers was used to verify that the pixel chips and their DAQ are working.
In the cosmic run 2022~\cite{koeppel_proceedings}, correlations between the pixel detector and the Mu3e scintillating fibre detector could be measured.
In addition, it was possible to reconstruct cosmic tracks with 4 hits in the vertex detector.
Configuration and data readout of the full pixel detector has been successfully performed by the DAQ throughout both campaigns.
These tests were only possible because of the helium cooling of the detector.
Without this efficient and light-weight cooling medium, only a fraction of pixel chips could have been operated without overheating.

\section{Discussion}

Cooling of a pixel tracking detector using gaseous helium under ambient conditions works, as this study demonstrates. 
Heat densities up to \qty{400}{\mWsqcm} were efficiently cooled. It is an excellent option for low-mass detector designs. Without the need for cooling pipes inside the active detector volume, the cooling merely does not contribute to the material a charged particle's trajectory experiences while passing the detector layers.

In an earlier report~\cite{MeierAeschbacher:2020ldo}, results from simulation were compared with measurements using a much simpler (and less accurate) mock-up. By switching to the miniature turbo compressor, we overcame the gas supply limitations when using compressed helium in a bottle. This enabled much more detailed studies, as shown, and we could demonstrate the concept of a continuously operating cooling plant. We did not find any significant deviation from expectations of the simulations. Also the hot region from the vortex (\cite[Fig.~11]{MeierAeschbacher:2020ldo}) was again present in our new mock-up studies. This strengthens our high level of trust into the flow simulations we have performed so far. We also measured the performance of the cooling under different heat loads and varying mass flows, confirming the linear behaviour expected from simulation.

However, one notable limitation exists.
While the heat-load densities are comparable to other pixel tracker detectors, the temperature gradients the pixel sensors see are much larger than the conventional designs using liquid or bi-phase cooling.
The pixel chips in the Mu3e experiment do not suffer as much from bulk damage as the sensors of other particle physics experiments because the radiation is mainly limited to an ionising dose. 
Hence operating temperatures up to near the glass transition temperatures of the adhesives used for the construction are acceptable. For applications where this is not acceptable, lower helium temperatures are possible, only limited by what can be reached using heat exchangers and the performance of the insulation of the piping needed to feed the helium to the detector. The gradient is still present but at a lower overall temperature.

We acknowledge certain limitations of our measurements.
A notable fraction of the mock-up detector had dead channels (chips that could not be heated or the temperature reading did fail). In part due to geopolitical circumstances, we were unable to improve on this in time of the project needs. The effect of these dead channels was estimated with Monte-Carlo techniques and we found the uncertainties to be well controlled and understood.

Miniature turbo compressors are a key ingredient to propel the helium. This allowed to reduce the cost by more than a factor of 10, compared to e.g.~oil-lubricated screw compressors. Their small form-factor allows for an easy integration into a rack. The helium cooling system was used by the Mu3e experimenters for weeks after our studies without any failure.
The components used in this study will be transferred to the final cooling plant for serving the circuit to cool the vertex pixel detector with \qty{2}{\gps} of helium. Upscaling this concept to \qty{16}{\gps} begun and will be installed in the forthcoming months in three copies for the benefit of the Mu3e collaboration to cool the much larger outer layer pixel detectors~\cite{Mu3e:2020gyw}.

This novel and innovative cooling approach is not limited to pixel detectors. Any active surface with comparable heat-load densities can be cooled efficiently using the technique we report here, as long as the helium flow can be guided appropriately.

\section[Author contributions]{Author contributions\footnote{Following CRediT (Contributor Roles Taxonomy), \url{https://credit.niso.org/}}}

\textbf{Thomas Theodor Rudzki:} \textit{(Sections 1, 2, 3)} Methodology, Software, Investigation, Formal analysis, Writing - Original Draft, Writing - Review \& Editing, Visualization
\textbf{Frank Meier Aeschbacher:} \textit{(Sections 1, 2, 3, 4)} Conceptualization, Methodology, Writing - Original Draft, Writing - Review \& Editing, Visualization, Supervision, Project administration 
\textbf{Marin Deflorin:} \textit{(Section 2.2)} Methodology, Investigation, Formal analysis, Visualization, Writing - Original Draft
\textbf{Niculin Flucher:} \textit{(Section 2.2)} Methodology, Investigation, Formal analysis, Writing - Original Draft

\section{Acknowledgements}
We would like to thank for the excellent support by the technicians and engineers at FHNW, Institute of Thermal and Fluid Engineering and the staff members at PSI.
We thank the Mu3e collaboration for all the effort that went into the successful integration and operation of the vertex detector, and the mechanical workshop of the Physics Institute Heidelberg whose great work made the presented measurements possible.
FMA would like to express thanks for the help and support in multiple discussions with Prof.~Beat Ribi of FHNW.
Without that excellent collaboration, nothing meaningful would have happened. 
We thank Celeroton AG for picking up on the challenge to create a special version of their compressors, optimised for the low-density regime of helium at ambient pressure. 
We gratefully acknowledge work done by our technician Michael Tröndle (for improvements to the rack setup), who is funded by Swiss National Science Foundation Project No. FLARE 20FL21\_201540.
TTR thanks the HighRR research training group [GRK 2058] for their support.
We thank LTU for their excellent product. 
We hope that peace will come soon.

\appendix

\section{Cooling parameters of various pixel barrel detectors}
\label{app:pixel_barrel_table}
The values of \autoref{tab:pixeldetscomparison} were calculated using the values listed below. We also give direct references to the documents where we took the values from.

\textit{CMS LHC Run 1:} BPix area: \qty{0.78}{\meter^2}, total pixel power: \qty{3.6}{\kilo\watt}, barrel area fraction: $\qty{0.78}{\meter^2} / (\qty{0.78}{\meter^2}+\qty{0.28}{\meter^2})$.~\cite[p.~34 \& 54]{CMS:2008xjf}

\textit{CMS Phase 1 Upgrade: } BPix area: 1184~$\textrm{modules} \times 16.4 \times \qty{64.8}{\milli\meter^2} = \qty{1.2}{\meter^2}$, max. BPix power: \qty{6}{\kilo\watt}.~\cite[p.~3, 6 \& 41]{CMSTrackerGroup:2020edz}

\textit{ATLAS LHC Run 1:} Area: 1456~$\textrm{modules} \times 24.4 \times \qty{63.4}{\milli\meter^2} = \qty{2.25}{\meter^2}$, total power: \qty{6}{\kilo\watt} (Front-end electronics power) + \qty{4}{\kilo\watt} (cables + regulators ID volume).\cite[p.~60-61, 82]{ATLAS:2008xda}

\textit{ALICE LHC Run 1:} Area layers 1-2: \qty{0.21}{\meter^2}, total power: \qty{1.35}{\kilo\watt}.~\cite[p.~3, 23]{ALICE:2008ngc}

\textit{ALICE Upgrade IB/OB:} IB area: 432 ~$\textrm{chips} \times 15 \times \qty{30}{\milli\meter^2} = \qty{0.19}{\meter^2}$, max. IB heat density: \qty{300}{\milli\watt/\centi\meter^2}; OB area: 23688 ~$\textrm{chips} \times 15 \times \qty{30}{\milli\meter^2} = \qty{10.7}{\meter^2}$, max. OB heat density: \qty{100}{\milli\watt/\centi\meter^2}.~\cite[p.~7, 61]{ALICE:2013nwm}

\textit{STAR:} Area: \qty{0.16}{\meter^2}~\cite{Schambach:2014uaa}, power dissipation: \qty{170}{\milli\watt/\centi\meter^2}~\cite{Greiner:2011zz}.

\textit{Belle II PXD:} Area of pixel matrix and switchers (cooled by gas): 16 ~$\textrm{chips} \times 15 \times \qty{44.8}{\milli\meter^2}$ + 24 ~$\textrm{chips} \times 15 \times \qty{61.44}{\milli\meter^2} = \qty{0.33}{\meter^2}$~\cite{Belle-IIDEPFET:2021pib}, power of pixel and switchers: 40 ~$\textrm{chips} \times \qty{0.5}{\watt} \times \qty{1}{\watt} = \qty{60}{\watt}$~\cite{Ye:2016rwb}.

\textit{Mu3e:} Area vertex:  108~$\textrm{chips} \times 20.66 \times \qty{23.18}{\milli\meter^2} = \qty{0.052}{\meter^2}$, area outer layers: 2736~$\textrm{chips} \times 20.66 \times \qty{23.18}{\milli\meter^2} = \qty{0.052}{\meter^2}$, heat density: \qty{250}{\milli\watt/\centi\meter^2}.~\cite[p.~21, 60]{Mu3e:2020gyw}

\bibliographystyle{elsarticle-num}
\bibliography{mu3eHeliumVertex}



\end{document}